\documentclass[fleqn,usenatbib]{mnras}

\usepackage{newtxtext,newtxmath}

\usepackage[T1]{fontenc}

\DeclareRobustCommand{\VAN}[3]{#2}
\let\VANthebibliography\thebibliography
\def\thebibliography{\DeclareRobustCommand{\VAN}[3]{##3}\VANthebibliography}

\usepackage{graphicx}	
\usepackage{amsmath}	
\usepackage{pdflscape}
\usepackage{array}
\usepackage{dblfloatfix}
\usepackage{float}
\usepackage{threeparttable}
\usepackage{textgreek}

\newcommand{\Ha}{H\textalpha}
\newcommand{\Hb}{H\textbeta}
\newcommand{\Hg}{H\textgamma}
\newcommand{\Hd}{H\textdelta}
\newcommand{\Ion}[2]{#1\,{\sc #2}}
\newcommand{\Line}[3]{#1\,{\sc #2}~\textlambda#3}
\newcommand{\kms}{\mbox{$\mathrm{km~s^{-1}}$}}
\newcommand{\he}[1] {He\,{\sc #1}}
\newcommand{\hel}[2] {He\,{\sc #1}~\textlambda#2}
\newcommand{\Msun}{\mbox{$\mathrm{M}_{\rm \odot}$}}
\newcommand{\Rsun}{\mbox{$\mathrm{R}_{\rm \odot}$}}
\newcommand{\obj}{V2487~Oph}

\title[The orbital period of V2487 Oph]{The orbital period of the recurrent nova V2487 Oph revealed}

\author[P. Rodríguez-Gil et al.]{Pablo Rodríguez-Gil$^{1,2}$\thanks{E-mail: prguez@iac.es (PRG)},
Jesús M. Corral-Santana$^{3}$,
N. Elías-Rosa$^{4,5}$,
Boris T. G\"ansicke$^{6}$,
\newauthor
Margarita Hernanz$^{5,7}$
and Gloria Sala$^{8,7}$
\\
$^{1}$Instituto de Astrofísica de Canarias, E-38205 La Laguna, Tenerife, Spain\\
$^{2}$Departamento de Astrofísica, Universidad de La Laguna, E-38206 La Laguna, Tenerife, Spain\\
$^{3}$European Southern Observatory, Alonso de Córdova 2107, Vitacura, Casilla 19001, Santiago de Chile, Chile\\
$^{4}$INAF -- Osservatorio Astronomico di Padova, Vicolo dell'Osservatorio 5, 35122, Padova, Italy\\
$^{5}$Institute of Space Sciences (ICE, CSIC), Campus UAB, Camí de Can Magrans s/n, 08193, Barcelona, Spain\\ 
$^{6}$Department of Physics, University of Warwick, Coventry CV4 7AL, UK\\
$^{7}$Institut d'Estudis Espacials de Catalunya (IEEC), Barcelona, Spain\\
$^{8}$Departament de Física EEBE, Universitat Politécnica de Catalunya (UPC), Barcelona, Spain
}

\date{Accepted 2023. Received 2023; in original form 2023}

\pubyear{2023}

\begin{document}
\label{firstpage}
\pagerange{\pageref{firstpage}--\pageref{lastpage}}
\maketitle
\setcounter{footnote}{0}

\begin{abstract}
We present the first reliable determination of the orbital period of the recurrent nova \obj\ (Nova Oph 1998). We derived a value of $0.753 \pm 0.016$~d ($18.1 \pm 0.4$~h) from the radial velocity curve of the intense \Line{He}{ii}{4686} emission line as detected in time-series X-shooter spectra. The orbital period is significantly shorter than earlier claims, but it makes \obj\ one of the longest period cataclysmic variables known. The spectrum of \obj\ is prolific in broad Balmer absorptions that resemble a white dwarf spectrum. However, we show that they come from the accretion disc viewed at low inclination. Although highly speculative, the analysis of the radial velocity curves provides a binary mass ratio $q \approx 0.16$ and a donor star mass $M_2 \approx 0.21$~\Msun, assuming the reported white dwarf mass $M_1 = 1.35$~\Msun. A subgiant M-type star is tentatively suggested as the donor star. We were lucky to inadvertently take some of the spectra when \obj\ was in a flare state. During the flare, we detected high-velocity emission in the Balmer and \hel{ii}{4686} lines exceeding $-2000$~\kms\ at close to orbital phase 0.4. Receding emission up to $1200$~\kms\ at about phase 0.3 is also observed. The similarities with the magnetic cataclysmic variables may point to magnetic accretion on to the white dwarf during the repeating flares.
\end{abstract}




\section{Introduction}

The orbital period of the Galactic nova \obj\ (Nova Ophiuchi 1998) has eluded detection until now. This is certainly not because of any known inherent difficulties of observation, but rather because \obj\ remains insufficiently explored.

The analysis of the X-ray spectrum reported in \cite{hernanz+sala02-1} pointed to a magnetic white dwarf in the binary, with the spectrum resembling those observed in cataclysmic variables (CVs) of the intermediate polar class. In addition, they found a rather high plasma temperature ($\ge 48$\,keV) in the hard X-ray spectrum, which suggests accretion onto a massive white dwarf.

After the 1998 nova event \citep{nakanoetal98-1}, the suggestion of \cite{hachisuetal02-1} that the object is a recurrent nova and the finding of a preceding eruption in 1900 by \cite{pagnottaetal09-1} using Harvard College Observatory photographic plates placed this very fast nova ($t_3 = 8$\,d, with $t_3$ the time to decline by 3 mag from the nova peak brightness) in the limelight again \cite[see][for more details]{schaefer10-1}. \cite*{schaeferetal22-1} communicated the finding of "superflares" in \textit{K2} data obtained in 2009. These are observed to recur daily and have a duration of about one hour, which they attributed to the reconnection of accretion disc magnetic field lines after being twisted and amplified due to the disc motion, in a similar way as in the Sun and flare stars.

Under the assumption that \obj\ is a member of the U~Sco subclass, an orbital period of about one day was suggested \citep{hachisuetal02-1,schaefer10-1}. In this paper, we report on VLT/X-shooter time-series spectroscopy of \obj\ that allowed us to measure its orbital period. Section~\ref{sec:obs} deals with the spectroscopic data and their reduction. Section~\ref{sec:avg_spec} describes the average spectrum of \obj\ and provides a list of the most prominent spectral lines. In Section~\ref{sec:Results}, we present and describe our results, and give our conclusions in Section~\ref{sec_Conclu}.  

\section{Observing data}
\label{sec:obs}
We collected spectra of \obj\ over the nights of 2019 June 7, 8 and 9 with the X-shooter échelle spectrograph \citep{vernetetal11-1} mounted on UT2 at the 8.2-m VLT in Cerro Paranal, Chile. A total of 210 spectra in nodding mode following the usual ABBA pattern were obtained per spectrograph arm with a nodding throw of 4 arcsec. On every night, we used a repeating observing sequence that starts with an acquisition image followed by two ABBA nod cycles. The slit widths were 1.0, 0.9 and 0.9\,arcsec for the ultraviolet-blue (UVB), visible (VIS) and near-infrared (NIR) arms, respectively, which provided approximate central spectral resolutions of 55, 34 and 53\,\kms. We used integration times of 300, 312 and 315 s on the first two nights, that were increased by 50\,s on the last night due to cloud patches. The spectral data were reduced with version v2.8.5 of the ESO \textsc{reflex} pipeline. Every individual spectrum used in the analysis presented below is the average of an ABBA cycle except for a few spectra coming from AB blocks at the end of the nightly allocated observing time.

\subsection{The X-shooter acquisition image light curve}
\label{sec:acq_imgs_LC}

\begin{figure}
   \centering
   \includegraphics[width=0.90\columnwidth]{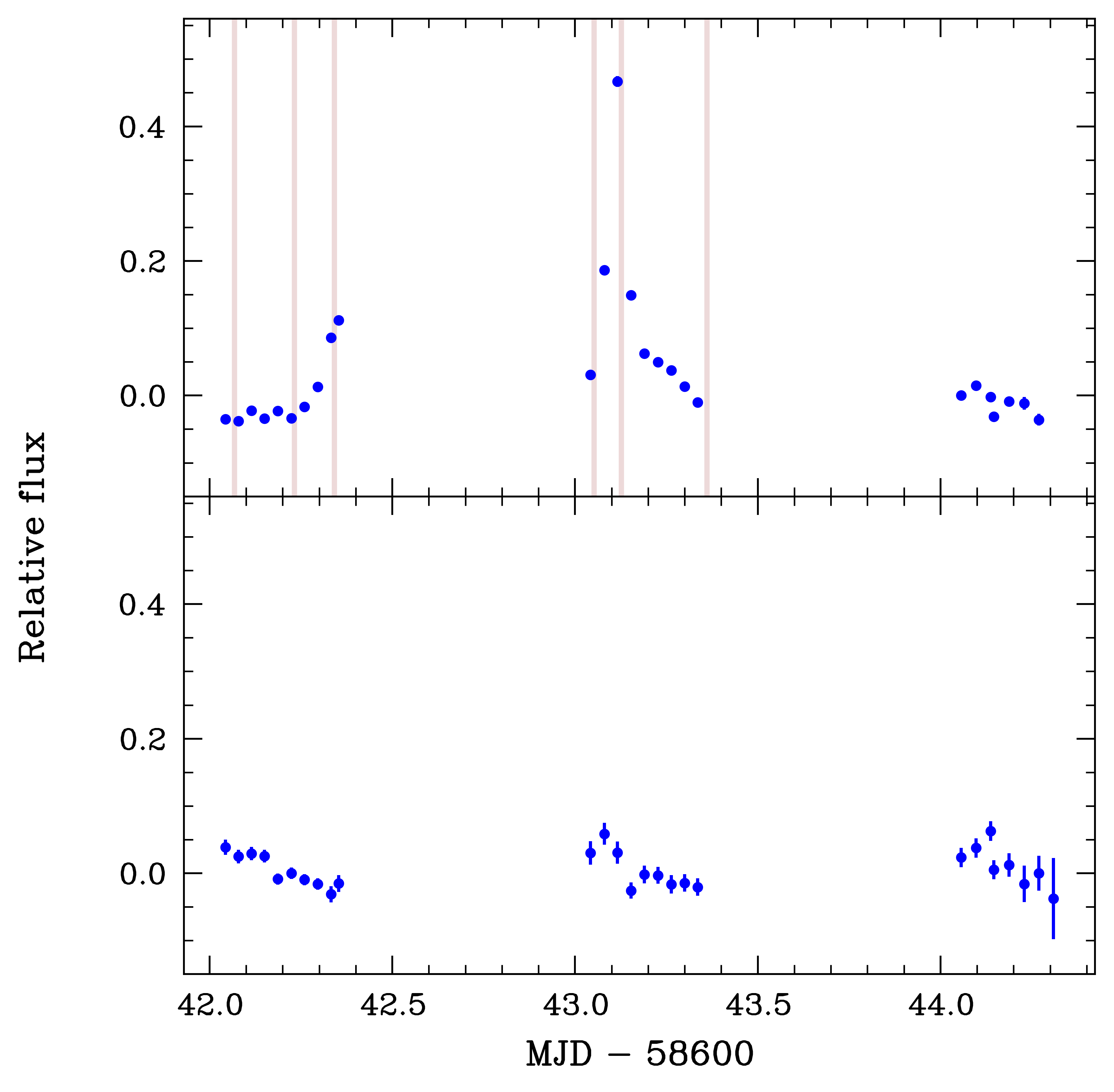}
    \caption{Top panel: light curve of \obj\ relative to a nearby comparison star derived using the X-shooter acquisition images for the three observing nights (2019 June 7, 8 and 9). A flare is detected at the beginning of the second night, while the onset of a flare may have been detected on the first night. The vertical lines mark the central times of the spectra shown in Fig.~\ref{fig:XSHOO_SPEC_FLARES}. Bottom panel: light curve of the comparison star used for \obj\ relative to a check star. The median flux of the whole data set has been subtracted in both panels for display purposes. 
    } 
	\label{fig:XSHOO_ACQ_LCs}
    \end{figure}

To check if any of our spectra were obtained by chance during a "superflare" of \obj\ (hereinafter flare), we first checked whether our X-shooter spectroscopy was unintentionally covered by any of \cite{schaeferetal22-1}'s light curves in which they detected flares. Unfortunately, the X-shooter spectra lie outside their time baseline. 
However, as explained in Section~\ref{sec:obs}, our observing strategy always started with an acquisition image, followed in general by two ABBA nod cycles. In order to check for flares in the acquisition images, we carried out variable aperture photometry of \obj\ and a comparison and a check star nearby with the HiPERCAM pipeline\footnote[1]{\url{https://cygnus.astro.warwick.ac.uk/phsaap/hipercam/docs/html/}}. The resulting light curves, shown in Fig.~\ref{fig:XSHOO_ACQ_LCs}, clearly show a flare on the second night and, possibly, the onset of another one on the first night. 



\section{The X-shooter spectrum of \obj}
\label{sec:avg_spec}

For the interstellar reddening correction we adopted a colour excess $E(B-V) = 0.6$, calculated using the 3D interstellar dust map of \cite{greenetal19-1}\footnote[2]{\url{http://argonaut.skymaps.info}} and the distance $d=6.4 \pm 1.6$\,kpc derived by \cite{bailer-jonesetal21-1} from the \textit{Gaia} eDR3 parallax. 

\begin{figure*}
   \centering
   \includegraphics[width=0.85\textwidth]{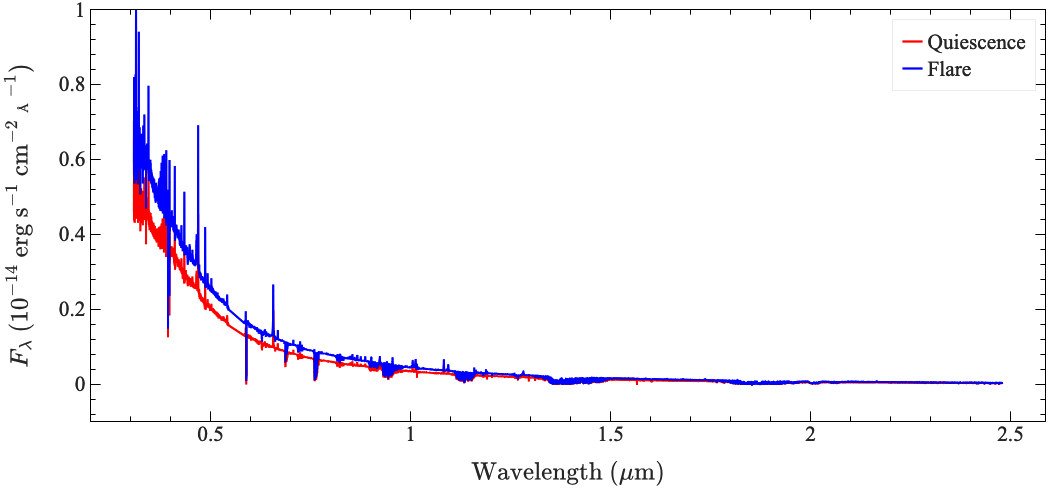}
    \caption{X-shooter average spectra of \obj\ in quiescence (red) and during a flare (blue). The fluxes were corrected for interstellar reddening using $E(B-V) = 0.6$. An online interactive version of this plot is available. Left-mouse-click on the legend label in the online plot to hide/show the sibling data set.} 
	\label{fig:avgspec}
    \end{figure*}

The unreddened average spectra (0.3080 to 2.4800~$\mu\mathrm{m}$) are presented in Fig.~\ref{fig:avgspec}. We illustrate in red the average of the spectra in quiescence, while in blue we plot the average of the spectra taken during a flare (i.e. the average of all the spectra acquired on the second night).

The spectrum of \obj\ displays a hot continuum consistent with the photometric spectral energy distribution presented in \cite{schaeferetal22-1}, which suggests a dominant contribution of the accretion disc. However, the H~\textsc{i} and some \he{i} emission lines are blended with much broader absorptions. At face value, these absorption lines might resemble those of a hot white dwarf. Adopting a white dwarf mass $M_\mathrm{WD}=1.35\,\mathrm{M}_\odot$ \citep{hachisuetal02-1}, $T_\mathrm{eff}=80\,000$\,K, and $d=6.4$\,kpc, the apparent magnitude of the white dwarf is $\simeq25$, i.e. any contribution of a white dwarf to the observed optical flux of \obj\ is negligible. This supports the conclusion that the Balmer \textit{absorption} lines originate in the accretion disc. In fact, as we will see in Section~\ref{sec:Res}, the phasing of the absorption-line radial velocity curves confirms this.   

The X-shooter spectrum of \obj\ also reveals several \Ion{He}{ii} emission lines, with the Fowler series \Line{He}{ii}{4686} line the strongest, exceeding \Hb\ in intensity. Remarkably, the spectrum shows a number of double-ionised oxygen emission lines, that are direct evidence for the Bowen fluorescence mechanism \citep{bowen34-1}. The \Ion{O}{iii} fluorescence lines involve trapping of \ion{He}{ii} Ly$\alpha$ $\lambda$303.782 photons by the \Ion{O}{iii} $2p^{\scriptscriptstyle 2}\,^{\scriptscriptstyle 3}\text{P}_{\scriptscriptstyle 2} -  2p3d\,^{\scriptscriptstyle 3}\text{P}_{\scriptscriptstyle 2}$ transition at 303.799~\AA, followed by cascade transitions (see e.g. \citet*{mcclintocketal75-1}; \citealp{schachteretal91-1}). Primary cascade  \Ion{O}{iii}~$\lambda\lambda3133, 3429, 3444$ 
(O1 channel) and secondary cascade \Ion{O}{iii}~$\lambda\lambda3299, 3312, 3341$ 
emissions are clearly detected in the UVB-arm spectrum of \obj, as is the \Line{O}{iii}{3122} fluorescence line 
as a result of excitation of the $2p3d\:^{\scriptscriptstyle 3}\text{P}_{\scriptscriptstyle 1}$ level by \ion{He}{ii} Ly$\alpha$ $\lambda$303.70 photons (O3 channel). The \Ion{C}{iii}/\Ion{N}{iii} $\lambda\lambda$4634--4651 Bowen fluorescence blend is also observed.

As we will later show, the radial velocity curves of the broad component in the most intense \Ion{O}{iii} emissions at 3133 and 3444~\AA\ indicate an origin on the white dwarf side of the binary system (Fig.~\ref{fig:RVC_OIII}). However, the narrow component follows the motion of the companion star. \Ion{O}{iii} fluorescence emission lines have also been reported in some strongly-magnetic CVs, known as polars. Examples are AM~Her, the precursor of the class \citep{raymondetal79-1,schachteretal91-1}, EF\,Eri \citep{schachteretal91-1}, and RX\,J1802.1+1804 \citep*{shraderetal97-1}. 
In Table~\ref{tab:line_ID} we list the wavelengths of the most prominent lines identified in the X-shooter spectrum of \obj.

\begin{table}
\caption[]{Most
prominent emission lines observed in the X-shooter average spectrum of \obj. Fluorescent lines are abbreviated as ``fluo.''}
\begin{center}
{\footnotesize
\begin{tabular}{lclclc}
& & & & &\\
\multicolumn{2}{l}{} &\multicolumn{2}{l}{}&\multicolumn{2}{l}{}\\
\hline\noalign{\smallskip}
{\bf Line} & $\lambda$ & {\bf Line} & $\lambda$& {\bf Line} & $\lambda$\\
& {($\mu$\bf{m})}& & {($\mu$\bf{m})}& & {($\mu$\bf{m})}\\
& & & & &\\ 
\hline\noalign{\smallskip}
\Ion{O}{iii}                 & 0.3122     &     \Ion{He}{i}                         & 0.4026 &    \Ion{Mg}{ii}        & 0.7896      \\
\Ion{O}{iii}                 & 0.3133 fluo. &     \Ion{O}{iii} + \Ion{C}{iii}         & 0.4071 &    \Ion{O}{i}          & 0.8446      \\
\Ion{He}{ii}                 & 0.3203     &     \Hd                           & 0.4102 &    Pa17                & 0.8467      \\
\Ion{O}{iii}                 & 0.3265     &     \Ion{He}{i}                         & 0.4121 &    \Ion{Ca}{ii} + Pa16 & 0.8498      \\
\Ion{O}{iii}                 & 0.3299 fluo. &     \Ion{C}{ii}                         & 0.4267 &    \Ion{Ca}{ii} + Pa15 & 0.8542      \\
\Ion{O}{iii}                 & 0.3312 fluo. &     \Hg                           & 0.4340 &    Pa14                & 0.8598      \\
\Ion{O}{iii}                 & 0.3341 fluo. &     \Ion{He}{i}                         & 0.4388 &    \Ion{Ca}{ii} + Pa13 & 0.8662      \\
\Ion{O}{iii}                 & 0.3429 fluo. &     \Ion{He}{i}                         & 0.4472 &    Pa12                & 0.8750	    \\
\Ion{O}{iii}                 & 0.3444 fluo. &     \Ion{Mg}{ii}                        & 0.4481 &    \Ion{Mg}{i}?        & 0.8807	    \\
\Ion{He}{i}                  & 0.3587     &     \Ion{C}{iii}/\Ion{N}{iii} fluo.       & Bowen  &    Pa11                & 0.8863	    \\
H18                          & 0.3692     &     \Ion{He}{ii}                        & 0.4686 &    Pa10                & 0.9015	    \\
H17                          & 0.3697     &     \Ion{He}{i}                         & 0.4713 &    Pa9                 & 0.9229	    \\
H16                          & 0.3704     &     \Hb                            & 0.4861 &    Pa\textepsilon     & 0.9546	    \\
H15                          & 0.3712     &     \Ion{He}{i}                         & 0.4922 &    Pa\textdelta          & 1.0049	    \\
H14                          & 0.3722     &     \Ion{He}{i}                         & 0.5016 &    \Ion{He}{ii}        & 1.0124	    \\
H13                          & 0.3734     &     \Ion{He}{ii}                        & 0.5412 &    \Ion{He}{i}         & 1.0830	    \\
\Ion{O}{iii}                 & 0.3760     &     \Ion{He}{i}                         & 0.5876 &    Pa\textgamma          & 1.0938      \\
H12                          & 0.3750     &     H$\alpha$                           & 0.6563 &    \Ion{He}{ii}        & 1.1626      \\
H11                          & 0.3771     &     \Ion{C}{ii}                         & 0.6578 &    Pa\textbeta           & 1.2818	    \\
H10                          & 0.3798     &     \Ion{C}{ii}                         & 0.6583 &    Br14                & 1.5880	    \\
\Ion{He}{i}                  & 0.3820     &     \Ion{He}{i}                         & 0.6678 &    Br12                & 1.6406	    \\
H9                           & 0.3835     &     \Ion{He}{i}                         & 0.7065 &    Br11                & 1.6806	    \\
H8                           & 0.3889     &     \Ion{He}{i}                         & 0.7281 &    \Ion{He}{i}         & 1.7002	    \\
\Ion{Ca}{ii}                 & 0.3934     &     \Ion{O}{i}                          & 0.7774 &    Br10                & 1.7361	    \\
H\textepsilon               & 0.3970     &     \Ion{Mg}{ii}                        & 0.7877 &    Br\textgamma          & 2.1660	    \\
\hline\noalign{\smallskip}
\end{tabular}
}

\end{center}

\label{tab:line_ID}
\end{table}
 
\section{Results}
\label{sec:Results}


   \begin{figure}
   \centering
   \includegraphics[width=1.01\columnwidth]{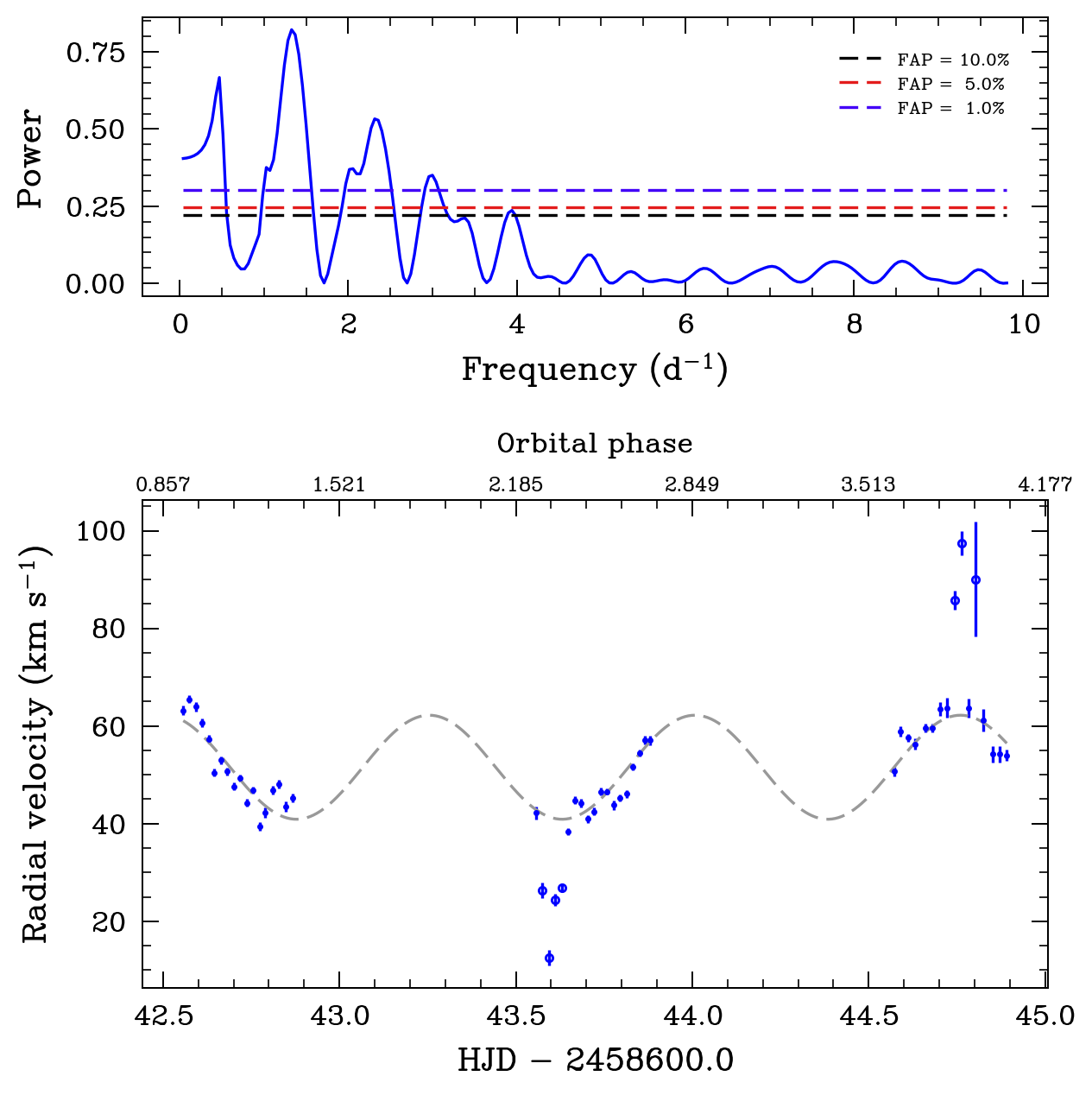}
    \caption{Top panel: Generalized Lomb-Scargle periodogram of the \hel{ii}{4686} radial velocity curve shown in the bottom panel. The horizontal dashed lines are the 10, 5, and 1 per cent false-alarm probability (FAP) levels. The highest peak provides a period of $0.753 \pm 0.016$~d ($18.1 \pm 0.4$~h). Bottom panel: radial velocity curve of the \hel{ii}{4686} emission line. We used a single Gaussian function with $\mathrm{FWHM} = 300$\,\kms\ as the cross-correlation template. Deviant points are marked as open circles and have been omitted from the period analysis and sinusoid fit (dashed line). The top x-axis shows the orbital phase computed using $T_0(\mathrm{HJD}) = 2\,458\,642.6076$.
    } 
	\label{v2487oph_mnras_fig03}
    \end{figure}

\subsection{The orbital period}\label{sec:Res}

We conducted a period search on the \hel{ii}{4686} emission line radial velocity curve, that was derived by cross-correlating the line profile in the average spectrum of every ABBA set with a $\mathrm{FWHM} = 300$\,\kms\ Gaussian template. Prior to this, the adjacent continuum was normalised to unity and the spectra were re-binned into a uniform velocity scale. 

We subjected the resulting radial velocity curve (bottom panel of Fig.~\ref{v2487oph_mnras_fig03}) to a generalised Lomb-Scargle analysis \citep*[GLS;][]{GLS09}, that uses a constant and a sine wave as the fitting function. We used the GLS class from \textsc{PyAstronomy}\footnote[3]{\url{https://github.com/sczesla/PyAstronomy}} \citep{pya}. The periodogram is illustrated in Fig.~\ref{v2487oph_mnras_fig03} (top panel). We also calculated the 10, 5, and 1~per cent false-alarm probability (FAP) levels in the frequency range shown. They should be interpreted as the probability that at least one out of all the independent power values in that range computed from a white Gaussian noise time series is as large or larger than a given periodogram power threshold. 

A period of $0.753 \pm 0.016$~d ($= 18.1 \pm 0.4$~h) is favoured, which we interpret as the orbital period of \obj. 



   \begin{figure}
   \centering
   \includegraphics[width=\columnwidth]{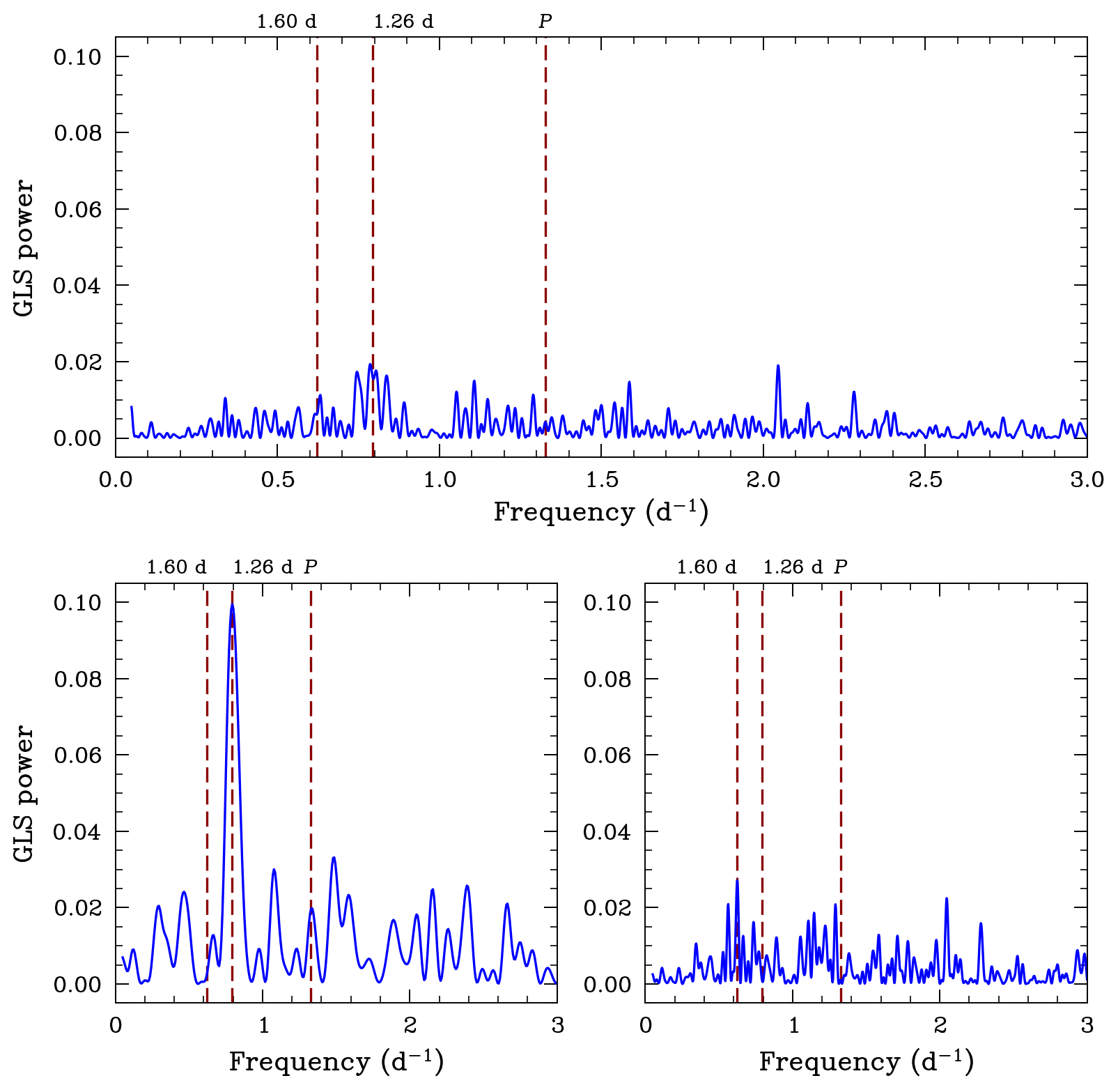}
    \caption{GLS periodograms of the \obj\ \textit{K2} light curve (data from table 3 in \citealt{schaeferetal22-1}). Top panel: periodogram using the whole \textit{K2} data set. Bottom left panel: same as the top panel, but using the first chunk of the \textit{K2} light curve (26-d time baseline). Bottom right panel: same as the top panel, but using the second chunk of the light curve (41-d time baseline). $P$ is the orbital period derived in this paper from the \Line{He}{ii}{4686} emission-line radial velocity curve (Section~\ref{sec:Res}). Note that the three plots share the same $y$-axis.
    } 
	\label{fig:GLS_K2}
    \end{figure}

\subsection{The \textit{K2} photometry revisited}

\cite{schaeferetal22-1} reported a tentative $1.24 \pm 0.02$\,d orbital period based on a Lomb-Scargle periodogram of their approximately 70-d long, 59-s integration \textit{K2} light curve from 2009 data. However, our value is about 40 per cent shorter. For this reason, we conducted a new period search on these \textit{K2} data, that they provide in their table~3. The results are presented in Fig.~\ref{fig:GLS_K2}. Both a GLS periodogram and a conditional entropy (CE; not shown) periodogram \citep{grahametal13-1} of the whole data set favour a 1.26-d periodicity. However, the periodograms seem to be dominated by the recurrence time scale of the flares in the first \textit{K2} light curve chunk, as illustrated by our periodogram computed from the first 26-d portion of the light curve. The GLS and CE periodograms calculated from the second chunk of the light curve only (41 d) suggest a different periodicity of 1.6~d, with little signal found at 1.26~d. Likewise, the 1.6-d signal is absent in the first chunk. In addition, there is no sign of the orbital period derived by us. This points to a variable flare recurrence time scale in \obj\ and makes a direct association with the orbital period or any clock in the system highly unlikely.    

\subsection{Quiescence and flare trailed spectra diagrams}

\begin{figure*}
   \centering
   \includegraphics[width=0.95\textwidth]{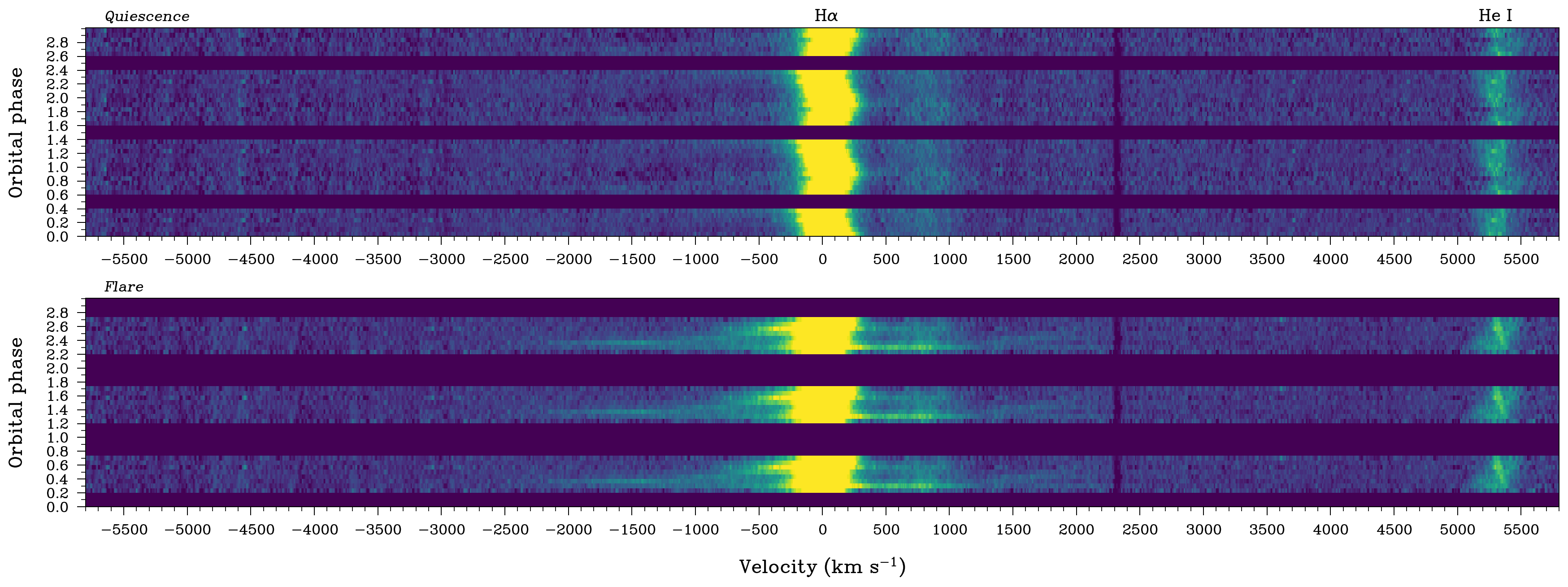}
    \caption{\Ha\ and \hel{i}{6678} trailed spectrograms in quiescence (top; data from the first and third nights) and during a flare (bottom; data from the second night). The spectra have been phase-binned into 15 orbital phase intervals and repeated twice for display purposes. The phase bins with no data are illustrated with dark horizontal stripes. 
    } 
   \label{fig:Ha_initial_trailed}
   \includegraphics[width=0.95\textwidth]{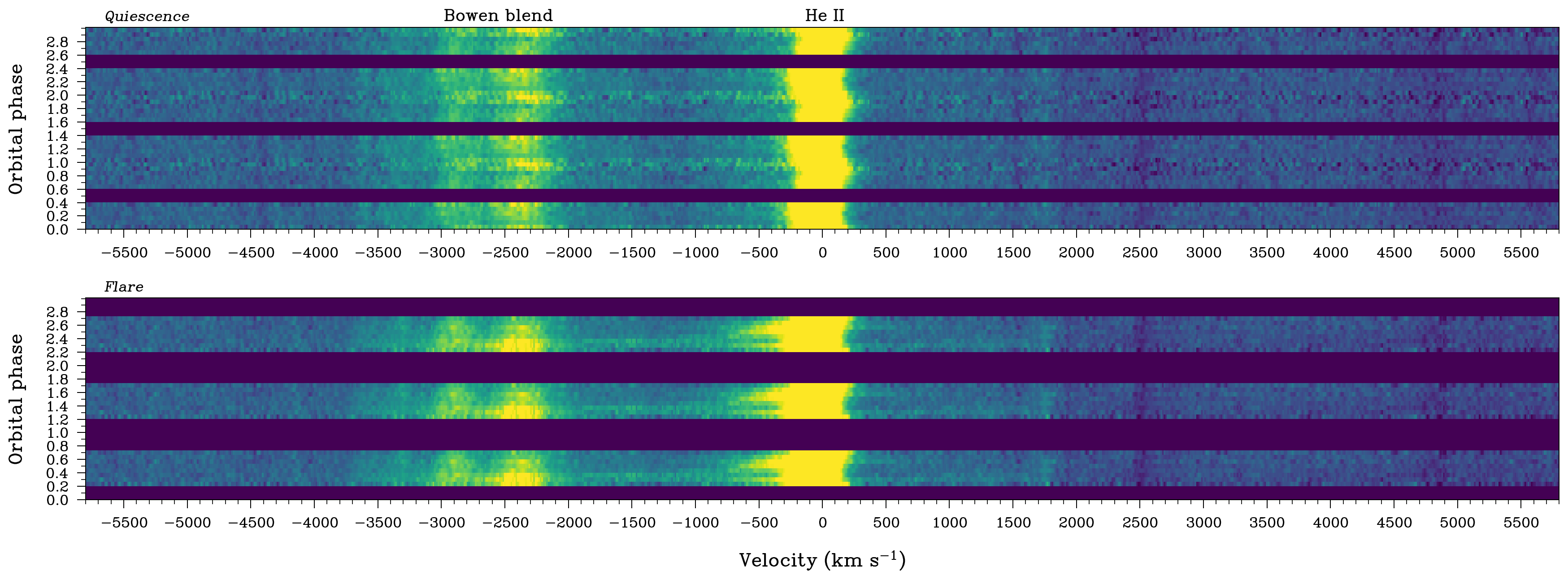}
   \caption{Same as Fig.~\ref{fig:Ha_initial_trailed}, but for the \Line{He}{ii}{4686} and Bowen blend emissions.}
   \label{fig:HeII4686_initial_trailed}
   \includegraphics[width=0.95\textwidth]{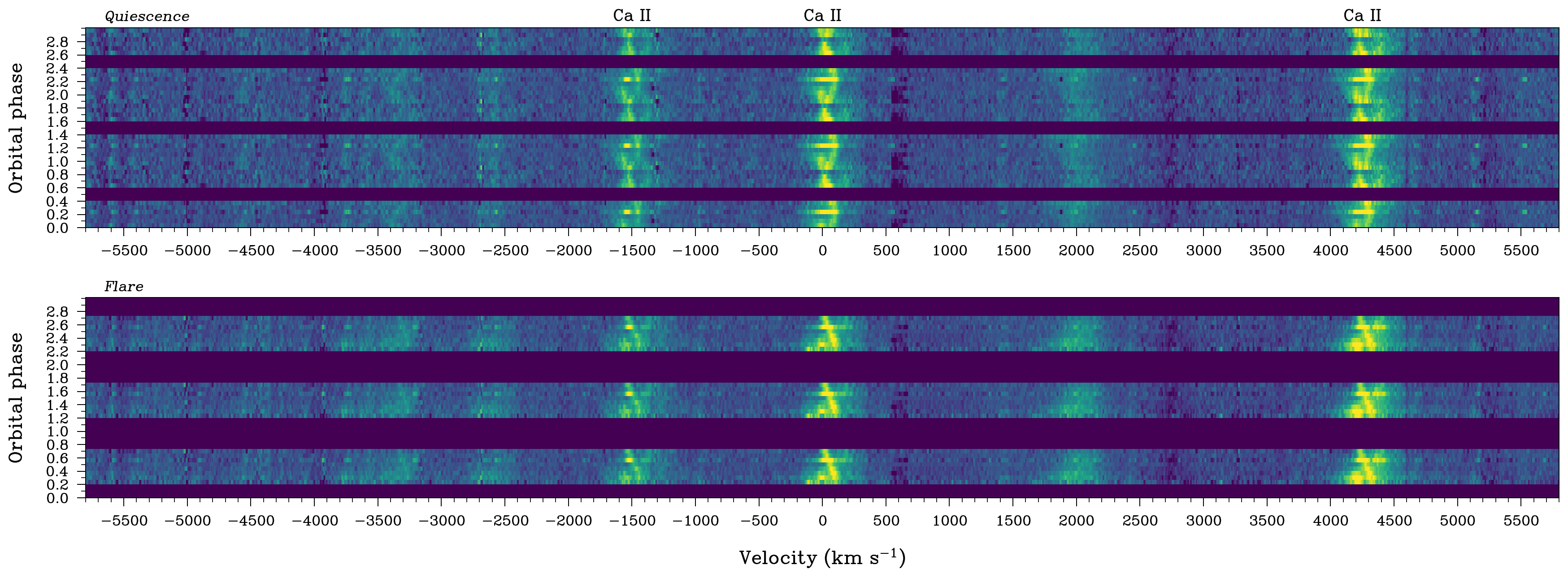}
   \caption{Same as Fig.~\ref{fig:Ha_initial_trailed}, but for the \Ion{Ca}{ii} emission triplet region.}
   \label{fig:CaII_initial_trailed}
    \end{figure*}

As described in Section~\ref{sec:acq_imgs_LC} (see also Fig.~\ref{fig:XSHOO_ACQ_LCs}), we covered a flare of the system on the second observing night. In order to test for changes between the quiescent and flare spectra with the orbital phase, we used the derived orbital period and the time of inferior conjunction of the companion star ($T_0$, calculated in Section~\ref{sec:irr_companion}) to assemble the trailed spectrograms that we present in Figs.~\ref{fig:Ha_initial_trailed}, \ref{fig:HeII4686_initial_trailed}, and \ref{fig:CaII_initial_trailed}. Despite the incomplete orbit coverage, there are evident differences.



The \Ha\ emission line is exemplary of the behaviour of the Balmer series emission. The trailed spectra during the flare (second night) show a ${\rm FWHM} \simeq 280$~\kms\ S-wave component with a maximum approaching radial velocity of $\simeq -1000$~\kms\ at about orbital phase $\varphi \simeq 0.4$. In addition, \Ha\ displays extra emission that extends to a very high approaching velocity, exceeding $-2000$~\kms, and is only visible in the close vicinity of $\varphi = 0.4$. High-velocity receding emission up to $1200$~\kms\ is also observed at $\varphi \simeq 0.3$. Resemblant high-velocity emission is detected in \Line{He}{ii}{4686} as well (see Fig.~\ref{fig:HeII4686_initial_trailed}), which also appears to show the extra emission peaking at $\simeq -2000$~\kms\ and probably the red counterpart.

The spectra from the second night cover the orbital phase range $\varphi = 0.26 - 0.69$, and the flare peak is observed at $\varphi = 0.35$ on that night. A curve of the normalized flux computed in the velocity range $(-2500,-1200)$~\kms\ relative to \Ha\ (not shown) reaches its maximum at $\varphi = 0.38$, which is consistent with the high-velocity emission being related to the flare.

\begin{figure*}
   \centering
   \includegraphics[width=0.95\textwidth]{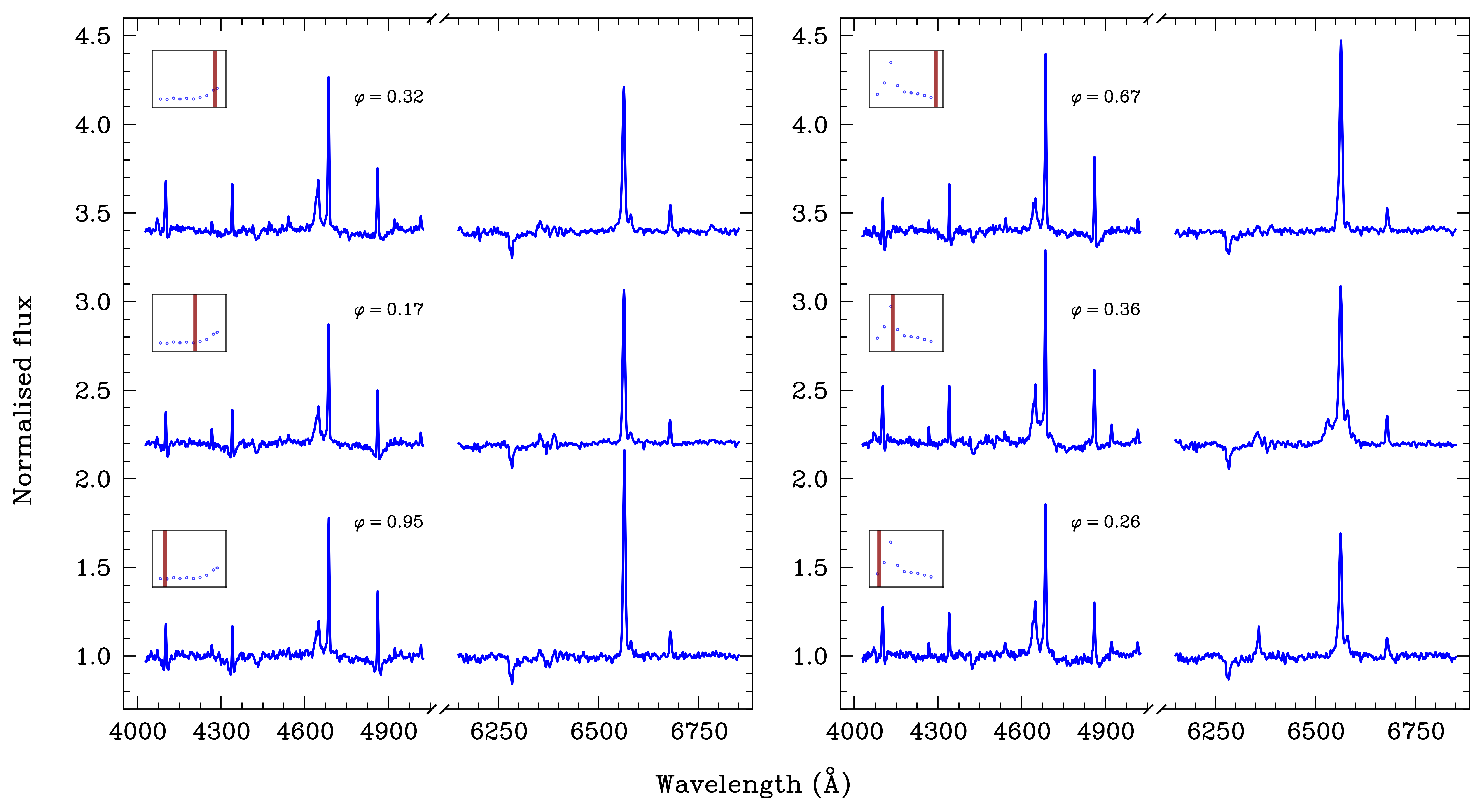}
    \caption{Changes in the spectrum of \obj\ with target brightness. The left and right panels correspond to the first and second nights, respectively. The insets show the central time of each spectrum in the light curve presented in Fig.~\ref{fig:XSHOO_ACQ_LCs}. Time runs from bottom to top.
    } 
	\label{fig:XSHOO_SPEC_FLARES}
    \end{figure*}

In Fig.~\ref{fig:XSHOO_SPEC_FLARES} we illustrate the observed line profile changes with the target brightness in selected spectral regions that contain Balmer and \ion{He}{i} lines, and the \Line{He}{ii}{4686} and Bowen blend emissions. The left and right panels of Fig.~\ref{fig:XSHOO_SPEC_FLARES} correspond to the first and second nights, respectively, and time runs from bottom to top.

On the first night, the \Hb, \Hg\ and \Hd\ broad absorptions originating in the accretion disc are noticeable in the first two spectra displayed ($\varphi = 0.95, 0.17$). However, they are filled up in the third one ($\varphi = 0.32$), which was obtained when \obj\ was at its brightest in our observing window, after what may be the onset of a flare. The same happens on the second night, with the accretion disc absorption lines only visible when the system was returning to quiescence after the flare ($\varphi = 0.67$ spectrum).

The changes in the emission line profiles are even more pronounced on the second night, when a flare peak was observed. At $\varphi = 0.36$, \Ha\ and \Line{He}{ii}{4686} display broad emission wings that reflect the presence of the high-velocity S-wave reaching its maximum blue-shift. This is also visible, albeit to a lesser extent, in \Hb\ and \Hg\ (see also Fig.~\ref{fig:trailed_UVB}). The \Line{He}{ii}{4686} and Bowen blend emissions are at their peaks at this orbital phase.

\cite{schaeferetal22-1} explained the flares in \obj\ as driven by reconnection of magnetic field lines above the accretion disc in a similar manner as in ordinary flare stars. However, it is unclear how this can be reconciled with our detection of high-velocity emission in the emission lines, that bears resemblance to magnetic accretion on to the white dwarf, as previously suggested by \cite{hernanz+sala02-1}. Before speculating on the flare-magnetic accretion link, more spectra with full orbital coverage during flares should be obtained to check, for example, whether the high-velocity emissions are observed at the same orbital phases as reported here.    

\subsection{Emission from the irradiated companion}\label{sec:irr_companion}

Besides the high-velocity emission during the flare, the trailed spectrogram in the vicinity of \Ha\ presented in Fig.~\ref{fig:Ha_initial_trailed} reveals a narrow emission S-wave in \Line{He}{i}{6678} with $\mathrm{FWHM} \simeq 50$\,\kms\ consistent with an origin on the companion star, possibly its heated face. The same emission component is seen in the \Ion{Ca}{ii} triplet during the flare and also in quiescence (Fig.~\ref{fig:CaII_initial_trailed}), as well as in the \Ion{O}{iii} fluorescence lines and \Ion{He}{i} (see e.g. the $\lambda5016$, $\lambda5876$, and $\lambda7065$ lines in Fig.~\ref{fig:trailed_VIS_01}).

\begin{figure}
   \centering
   \includegraphics[width=0.91\columnwidth]{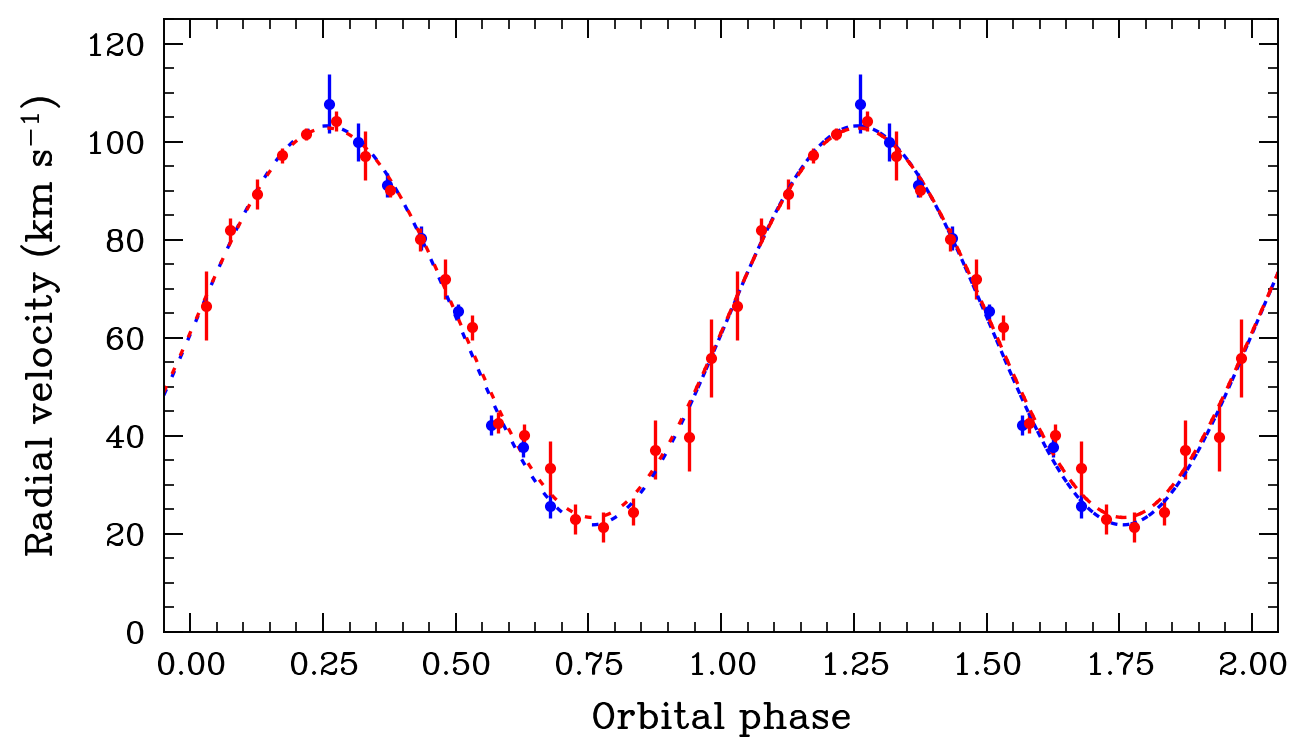}
    \caption{\Line{Ca}{ii}{8542} narrow emission radial velocity curves of the flare spectra (second night only; blue dots) and of all the spectra combined (red dots). The flare spectra were orbital phase-averaged into 15 bins, and the whole data set into 20 bins prior to measuring the radial velocities by carefully eyeballing the narrow emission component in the trailed spectra and using single Gaussian fitting. The dashed lines are the corresponding best sine fits. A full orbital cycle has been repeated for display purposes.  
    } 
	\label{fig:CaII8542_FLARE_vs_QUIES_RVEL}
    \end{figure}

In order to check whether the quiescence and flare spectra can be combined to improve orbital coverage, we used single Gaussian fitting to obtain the radial velocity curve of the \Line{Ca}{ii}{8542} narrow emission component during the flare (second night spectra only) and of all the spectra combined. The result is illustrated in Fig.~\ref{fig:CaII8542_FLARE_vs_QUIES_RVEL}, and shows that both radial velocity curves are consistent with each other. This way we achieved full-orbit coverage of this line.

Under the assumption that the narrow emission has its origin on the heated face of the companion star, we were able to obtain the time of its inferior conjunction,
$$T_0(\mathrm{HJD}) = 2\,458\,642.6076 \pm 0.019~.$$

\begin{figure}
   \centering
   \includegraphics[width=\columnwidth]{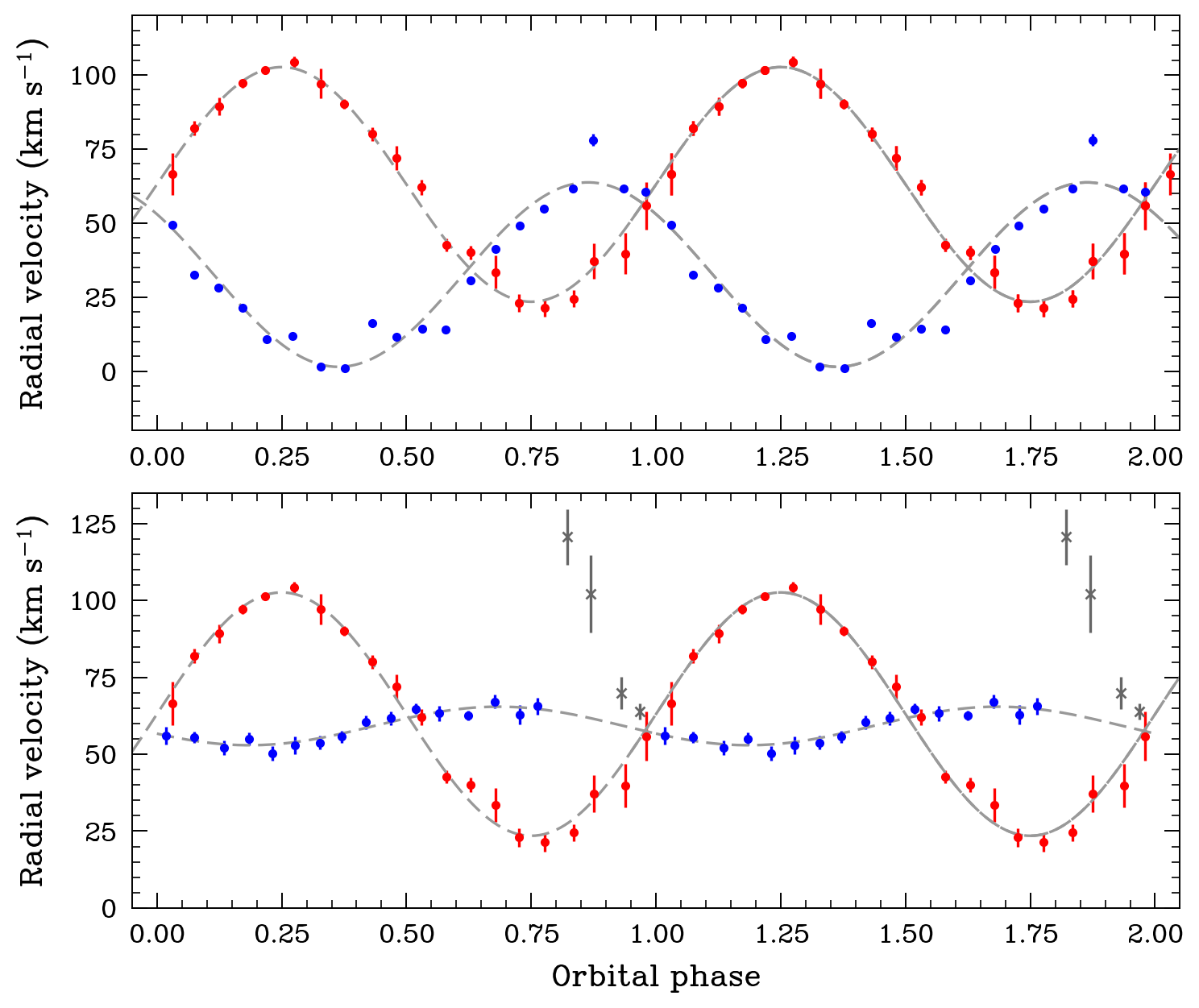}
    \caption{Top panel: \Line{Ca}{ii}{8542} (red dots) and \Line{He}{ii}{4686} (blue dots) emission-line radial velocity curves measured from the phase-binned spectra (20 bins). Assuming that the \Line{Ca}{ii}{8542} narrow emission S-wave follows the motion of the companion star, then \Line{He}{ii}{4686} is delayed by 0.12 cycle relative to the radial velocity curve expected for the white dwarf. Bottom panel: Phase-binned radial velocity curve through cross-correlation of the H\textepsilon--H10 disc absorption profiles with their average spectrum as the template (blue dots). The emission lines superimposed on the broad absorptions were masked out. The points plotted as grey crosses were excluded from the fitting procedure. The red dots are the same \Line{Ca}{ii}{8542} narrow emission-line radial velocity curve as the top panel. All data are plotted twice for continuity and the grey dashed lines are the respective best-fit sinusoids.} 
	\label{v2487oph_mnras_fig04}
    \end{figure}

The phase-binned \Line{Ca}{ii}{8542} narrow emission and the \Line{He}{ii}{4686} radial velocity curves are displayed in Fig.~\ref{v2487oph_mnras_fig04}. If we assume that the \Ion{Ca}{ii} narrow emission is representative of the orbital motion of the companion star, then the \Line{He}{ii}{4686} radial velocities are delayed by 0.12~cycle relative to the expected motion of the white dwarf, with a maximum blue excursion at phase 0.37. This phase offset is an ubiquitous characteristic of the SW~Sex stars and magnetic CVs \citep[e.g.][]{rodriguez-giletal07-2,rodriguez-giletal12-1,beuermannetal21-1}. This lends support to our assumption of the \Ion{Ca}{ii} narrow emission having an origin on the hemisphere of the companion star facing the white dwarf, and strenghtens the suggestion of \cite{hernanz+sala02-1} of magnetic accretion.

\begin{figure}
   \centering
   \includegraphics[width=\columnwidth]{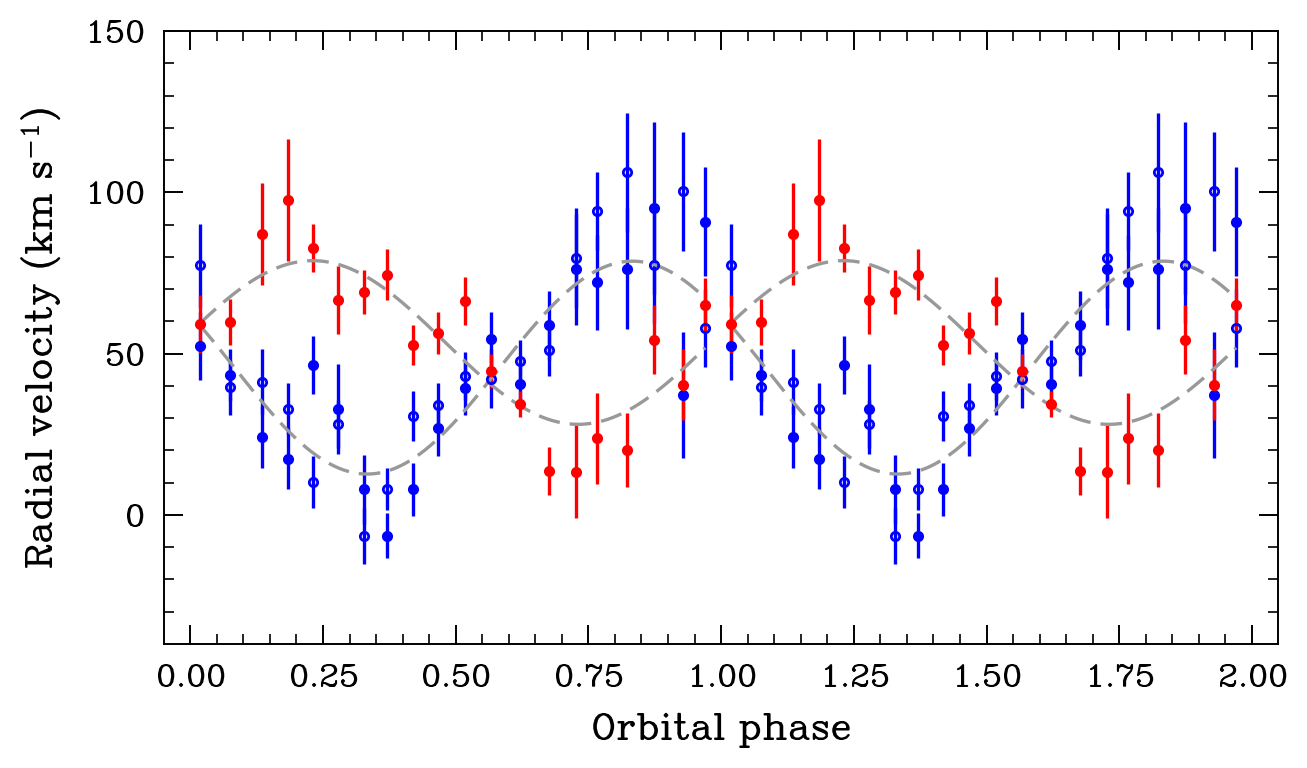}
    \caption{Phase-binned emission-line radial velocity curves of the broader component in \Line{O}{iii}{3133} (solid blue dots) and $\lambda$3444 (open blue circles) from cross-correlation with a $\mathrm{FWHM} = 300$\,\kms\ Gaussian template. For the narrow emission in \Line{O}{iii}{3444} (red dots) we used single Gaussian fitting. All data are plotted twice for continuity, and the best-fit sinusoids are superimposed. 
    } 
	\label{fig:RVC_OIII}
    \end{figure}

The radial velocity curves of the \Ion{O}{iii} $\lambda$3133 and $\lambda$3444 broad and narrow emissions are also offset relative to each other, as Fig.~\ref{fig:RVC_OIII} and Table~\ref{tab:velfits} illustrate, which also points to the companion star as the likely source for the \Ion{O}{iii} fluorescence narrow emission (see also Fig.~\ref{fig:trailed_NUV}). 

In Table~\ref{tab:velfits} we list the parameters of the sine fits to the emission-line radial velocity curves derived from the phase-binned spectra of the form:
\begin{equation}
V(\varphi) = \gamma + K \sin\left[2 \pi (\varphi - \varphi_0)\right]~,
\end{equation}
\noindent
with $V(\varphi)$ the radial velocity, $\gamma$ the systemic velocity, $K$ the radial velocity amplitude, and $\varphi_0$ the phase offset. To derive the radial velocities we used cross-correlation with a $\mathrm{FWHM} = 300$\,\kms\ Gaussian template except for the narrow emissions, for which we used single Gaussian fitting. Also note that we phase-binned the spectra from the three nights together.  


\begin{table}
\caption{Phase-binned radial velocity curve best-fit parameters.}
\begin{threeparttable}
\setlength{\tabcolsep}{0.95ex}
\label{tab:velfits}
\vspace*{-2.5ex}

\begin{tabular}[t]{lccc}
\hline\noalign{\smallskip}
Emission line &  $\gamma$ ~~~~& $K$~~~~ & $\varphi_0$~~~~\\
 &  (\kms)  ~~~~      & (\kms) ~~~~ & ~~~~\\    
\hline\noalign{\smallskip}
\smallskip
\Line{Ca}{ii}{8542} narrow             & $63.0 \pm 0.6$ ~~~~& $39.7 \pm 0.8$ ~~~~  & --- ~~~~\\
\Line{He}{i}{6678}             & $50.8 \pm 0.7$ ~~~~& $-32.1 \pm 0.9$ ~~~~  & $-0.058 \pm 0.005$ ~~~~\\
\Line{He}{ii}{4686}             & $31.6 \pm 0.1$ ~~~~& $-28.9 \pm 0.2$ ~~~~  & $0.116 \pm 0.001$  ~~~~\\
\Ha    & $34.1 \pm 0.1$ ~~~~& $-33.9 \pm 0.2$ ~~~~ & $0.148 \pm 0.001$ ~~~~\\
\Hb    & $28.6 \pm 0.2$ ~~~~& $-23.1 \pm 0.4$ ~~~~ & $0.086 \pm 0.002$ ~~~~\\
\Hg    & $39.5 \pm 0.4$ ~~~~& $-20.5 \pm 0.6$ ~~~~ & $0.045 \pm 0.004$ ~~~~\\
\Hd    & $39.4 \pm 0.6$ ~~~~& $-19.4 \pm 0.8$ ~~~~ & $0.015 \pm 0.007$ ~~~~\\
\noalign{\medskip}
\Ion{O}{iii} fluo.    & $44.6 \pm 1.8$ ~~~~& $-36 \pm 3$ ~~~~ & $0.076 \pm 0.010$ ~~~~\\
\Line{O}{iii}{3444} narrow    & $53 \pm 2$ ~~~~& $25 \pm 3$ ~~~~ & $-0.02 \pm 0.02$ ~~~~\\
\noalign{\medskip}
H\textepsilon--H10 absorp.\tnote{1*}    & $59.2 \pm 0.6$ ~~~~& $-6.3 \pm 0.7$ ~~~~ & $-0.06 \pm 0.02$ ~~~~\\
\noalign{\smallskip}\hline
\end{tabular}
\smallskip
\begin{tablenotes}\footnotesize
\item[1] Superimposed emission lines masked out.
\item[*] See Section~\ref{sec_ABS_RVCs}.
\end{tablenotes}
\end{threeparttable}
\end{table}


\subsection{Radial velocity curve of the accretion disc absorption lines}\label{sec_ABS_RVCs}

In Section~\ref{sec:avg_spec}, we showed that the contribution of the white dwarf to the optical flux of \obj\ is negligible. Thus, the broad absorption lines detected in the spectrum are very likely coming from an accretion  disc observed at low inclination. To test this, we measured the radial velocities of the Balmer absorption lines by cross-correlation of the H\textepsilon--H10 profiles with their average spectrum as template, after masking the superimposed emission profiles out. The resulting radial velocity curve is presented in the bottom panel of Fig.~\ref{v2487oph_mnras_fig04}. The phasing, with the minimum and maximum velocity at $\varphi \simeq 0.25$ and $\varphi \simeq 0.75$, respectively, confirms that the observed absorption lines likely originate in the accretion disc. In simpler terms, by using the radial velocity curve of the \Ion{Ca}{ii} emission in conjunction with this information, we can make an educated guess about the binary mass ratio of \obj. The velocity amplitude of the \Line{Ca}{ii}{8542} narrow emission $K_\mathrm{em} = 39.7$~\kms provides an estimate of the companion star radial velocity amplitude $K_2$. On the other hand, let us assume that the amplitude of the H\textepsilon--H10 absorptions radial velocity curve reflects that of the white dwarf, so $K_1 = 6.3$~\kms\ (see Table~\ref{tab:velfits}). This provides a binary mass ratio $q = M_2/M_1 = K_1/K_2  = 0.16$. Adopting $M_1 = 1.35$~\Msun\ (\href{http://arxiv.org/abs/astro-ph/0110265}{Hachisu et al.~2002}), the mass of the companion star would then be $M_2 = 0.21$~\Msun\ and its equivalent Roche lobe radius $R_\mathrm{L_2} = 0.96$\,\Rsun. For a Roche-lobe filling companion star, this might suggest an evolved M-type star in \obj, possibly a subgiant, considering the orbital period of 18.1\,h, as suggested by \cite{darnleyetal12-1}. However, the H\textepsilon--H10 absorption lines are contaminated with narrow emission lines from the companion star that can introduce unwanted systematics even if masked to derive their radial velocity curve via cross-correlation with their average profile, not to mention that \obj\ is very likely an intermediate polar, so vertical velocity components cannot be neglected. Thus, we deem our assumption that the radial velocity amplitude of these absorptions traces the motion of the white dwarf dubious.

The very hot continuum in \obj\ shows the dominance of the accretion flow brightness, as already noted by \cite{schaeferetal22-1}, leaving very little chance of detecting any spectral features of the companion star. However, in order to maximize our chances of detecting any, we eyeballed the trailed spectra diagrams constructed using the quiescence spectra only (15 phase bins; Figs.~\ref{fig:trailed_QUIES_UVB_01} and~\ref{fig:trailed_QUIES_VIS_01}) to search for absorption lines that may reflect its motion, i.e. radial velocity maxima at orbital phases 0.25 (red) and 0.75 (blue).

We observe candidate absorption lines from the companion star at approximate wavelengths 4148, 4572, 7040, 7480, 7940, 7948, 8088, and 8430~\AA, with the last two broader than the rest. There is a strong \Ion{Fe}{i} line in solar-type stars at 4148~\AA, that remains detectable in a wide range of temperature. The possible absorption at 4572~\AA\ could be \Ion{Ti}{ii}. It is present in warmer stars and disappears in late G and K stars. At 7948~\AA, a \Ion{Rb}{i} line is seen in stars with solar temperature or cooler. Note that this is quite speculative and inconsistent in terms of spectral classification, and casts doubts on the tentative donor star mass derived above. However, it hints at the presence of the companion star spectrum and indicates the need for deeper spectra, something feasible for a long orbital period CV such as \obj\ while keeping a proper sampling of the binary orbit.


\section{Conclusions}\label{sec_Conclu}
We have measured the orbital period of \obj\ to be $0.753 \pm 0.016$~d ($18.1 \pm 0.4$~h) from a radial velocity study of its intense \hel{ii}{4686} emission line. Our value is about 40 per cent shorter than the 1.24\,d suggested by the analysis of the \textit{K2} light curve presented in \cite{schaeferetal22-1}. We show that no significant power at 0.753\,d is present in the periodogram of the whole \textit{K2} photometry data set. 

We found that the broad Balmer absorption lines seen in quiescence (i.e. out of flare) are produced in the accretion disc. At the distance of \obj\ (6.4~kpc), any contribution of the white dwarf to the optical flux can be neglected.

Narrow emissions originating on the heated face of the companion star are observed in the \Ion{O}{iii} fluorescence lines, \Ion{He}{ii}, and the \Ion{Ca}{ii} triplet in the flare spectra. Irradiation may also be present in quiescence. 

The trailed spectrograms in quiescence may show absorption lines from the companion star, but this needs further investigation now that the orbital period of \obj\ is known and longer exposure spectra can be taken while still properly sampling the binary orbit.

Analysis of the radial velocity curves resulted in a binary mass ratio $q \approx 0.16$. Assuming a white dwarf mass $M_1 = 1.35$~\Msun\ \citep{hachisuetal02-1}, a possibly M-type subgiant donor star with mass $M_2 \approx 0.21$~\Msun\ is very tentatively suggested.

The X-shooter spectra from the second night were serendipitously obtained when \obj\ was experiencing a flare, during which high-velocity Balmer and \hel{ii}{4686} emissions exceeding $-2000$~\kms\ are observed in the blue wings at about $\varphi = 0.4$. 

Receding emission is also seen extending up to $1200$~\kms\ at $\varphi \simeq 0.3$. We suggest that the high-velocity emissions have to do with magnetic accretion on to the white dwarf, as observed in magnetic CVs.

To gain further insight into the spectral changes during the flares relative to quiescence, time-resolved spectroscopy on several nights that provide full orbital coverage should be obtained before the flaring activity in \obj\ stops. This would also provide phase-binned spectra in quiescence with a better signal-to-noise ratio to test for absorption lines from the companion star and hopefully reveal its spectral type.     

\section*{Acknowledgements}

The authors thank Mike Shara for reviewing the manuscript and for his encouraging report. We also thank Carlos Allende for useful discussion about the candidate absorption features from the companion star. We acknowledge valuable discussion with Brad Schaefer. Based on observations collected at the European Organisation for Astronomical Research in the Southern Hemisphere under ESO programme 103.D--0740(A). PR-G acknowledges support from the Consejería de Economía, Conocimiento y Empleo del Gobierno de Canarias and the European Regional Development Fund (ERDF) under grant with reference ProID2021010132 and ProID2020010104, and also acknowledges support from the ESO Scientific
Visitor Programme in Vitacura, Chile. NE-R acknowledges partial support from MIUR, PRIN 2017 (grant 20179ZF5KS), from PRIN-INAF 2022 and from the Spanish MICIN/AEI grant PID2019-108709GB-I00. This project has received funding from the European Research Council (ERC) under the European Union’s Horizon 2020 research and innovation
programme (Grant agreement No. 101020057). MH acknowledges funding support from the MICIN/AEI grant PID2019-108709GB-I00, EU-FEDER funds and the AGAUR/Generalitat de Catalunya grant SGR-01526/2021. GS acknowledges funding support from the MICIN/AEI grant PID2020-117252GB-I00, EU-FEDER funds and the AGAUR/Generalitat de Catalunya grant SGR-386/2021. This research was supported in part by the National Science Foundation under Grant No. NSF PHY-1748958.

\section*{Data Availability}
 
The data underlying this article will be shared on reasonable request to the corresponding author.



\bibliographystyle{mnras}
\bibliography{v2487oph_porb_mnras} 




\appendix

\section{Trailed spectra plots}
\begin{figure*}
\begin{center}
\includegraphics[width=\linewidth]{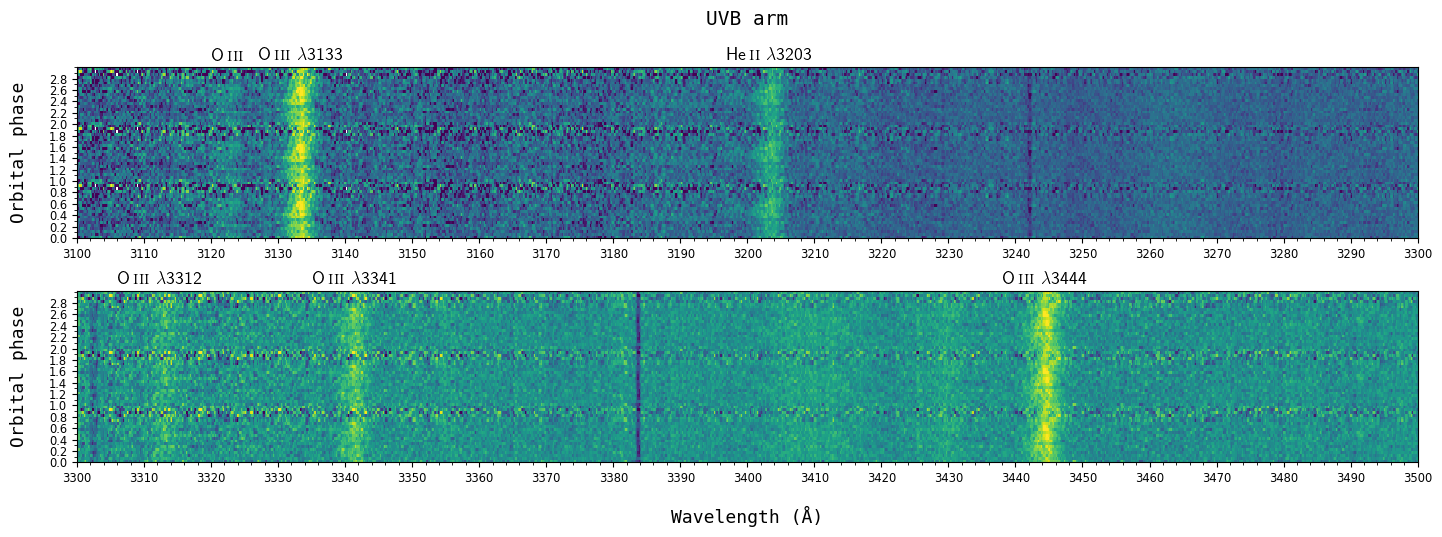}
\caption{Trailed spectra diagrams in the far blue fluorescence emission line region after averaging all the spectra into 20 phase bins. The orbital cycle is repeated twice. The two most intense \Ion{O}{iii} emission lines are labelled.}
\label{fig:trailed_NUV}
\end{center}
\end{figure*}

\begin{landscape}
\begin{figure}
\begin{center}
\includegraphics[width=0.9\linewidth]{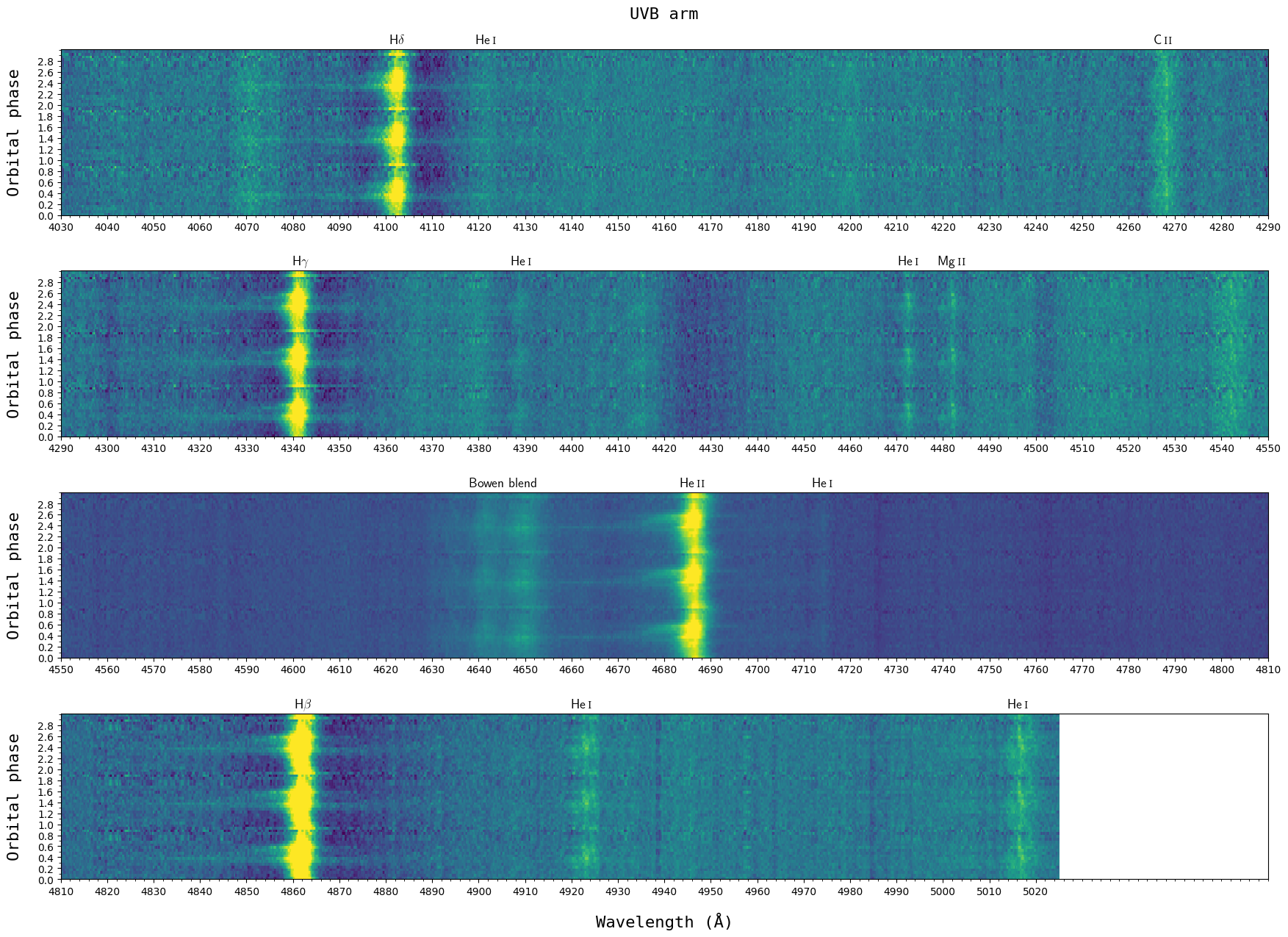}
\caption{X-shooter UVB arm trailed spectra diagrams using all the data averaged into 20 orbital phase bins. The full cycle has been repeated twice for clarity.}
\label{fig:trailed_UVB}
\end{center}
\end{figure}
\end{landscape}

\begin{landscape}
\begin{figure}
\begin{center}
\includegraphics[width=0.9\linewidth]{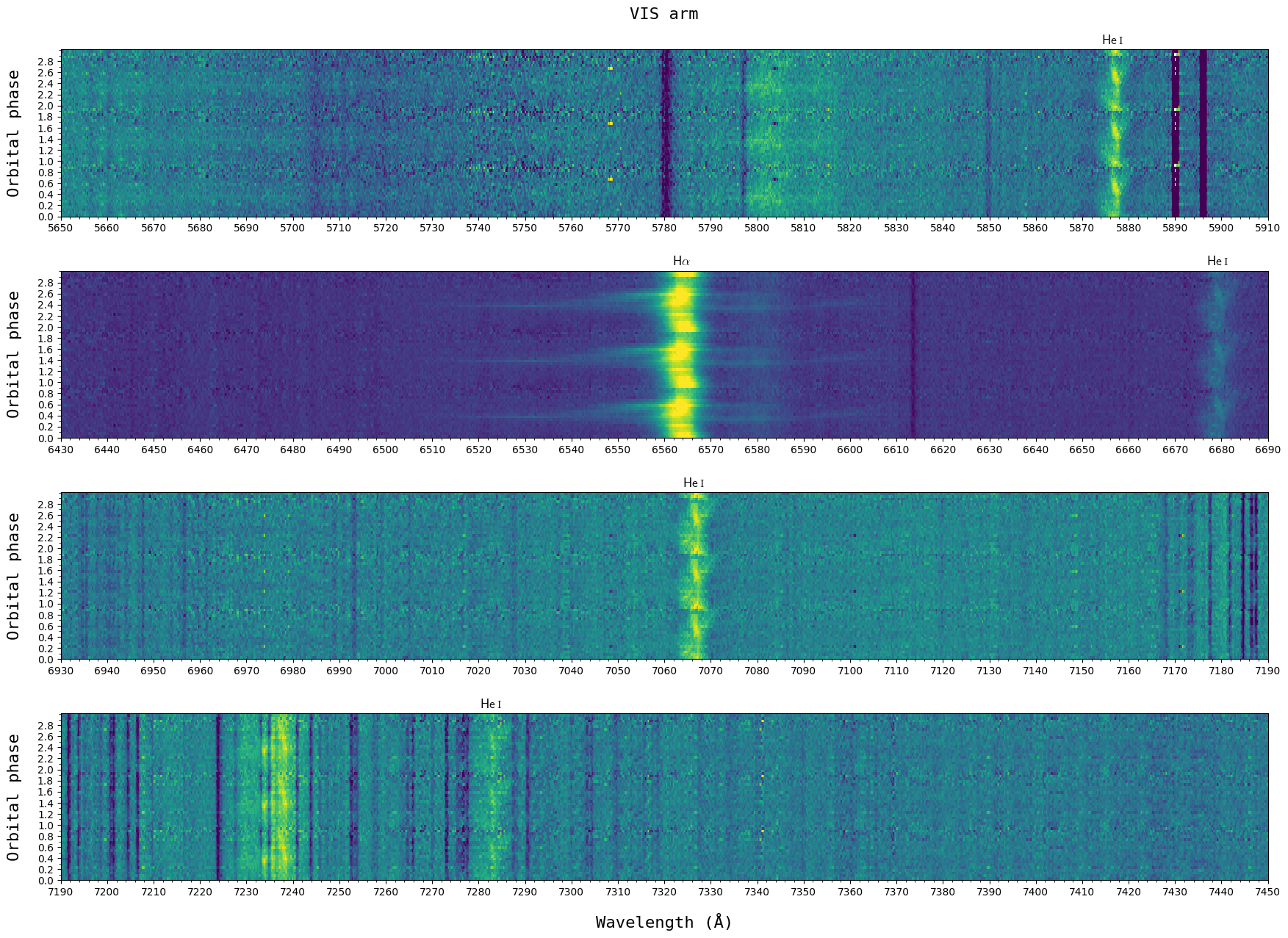}
\caption{Same as Fig.~\ref{fig:trailed_UVB}, but for the VIS arm.}
\label{fig:trailed_VIS_01}
\end{center}
\end{figure}
\end{landscape}

Figs.~\ref{fig:trailed_NUV} to \ref{fig:trailed_VIS_01} illustrate the trailed spectrograms using all the spectra (quiescence and flare state) phase-binned into 20 bins, while the trailed spectra plots presented in Figs.~\ref{fig:trailed_QUIES_UVB_01} to \ref{fig:trailed_QUIES_VIS_01} were constructed using the quiescence spectra only (first and third night) after applying a 15-bin phase binning.



\begin{landscape}
\renewcommand{\thefigure}{\thesection\arabic{figure} (cont.)}
\addtocounter{figure}{-1}
\begin{figure}
\begin{center}
\includegraphics[width=0.9\linewidth]{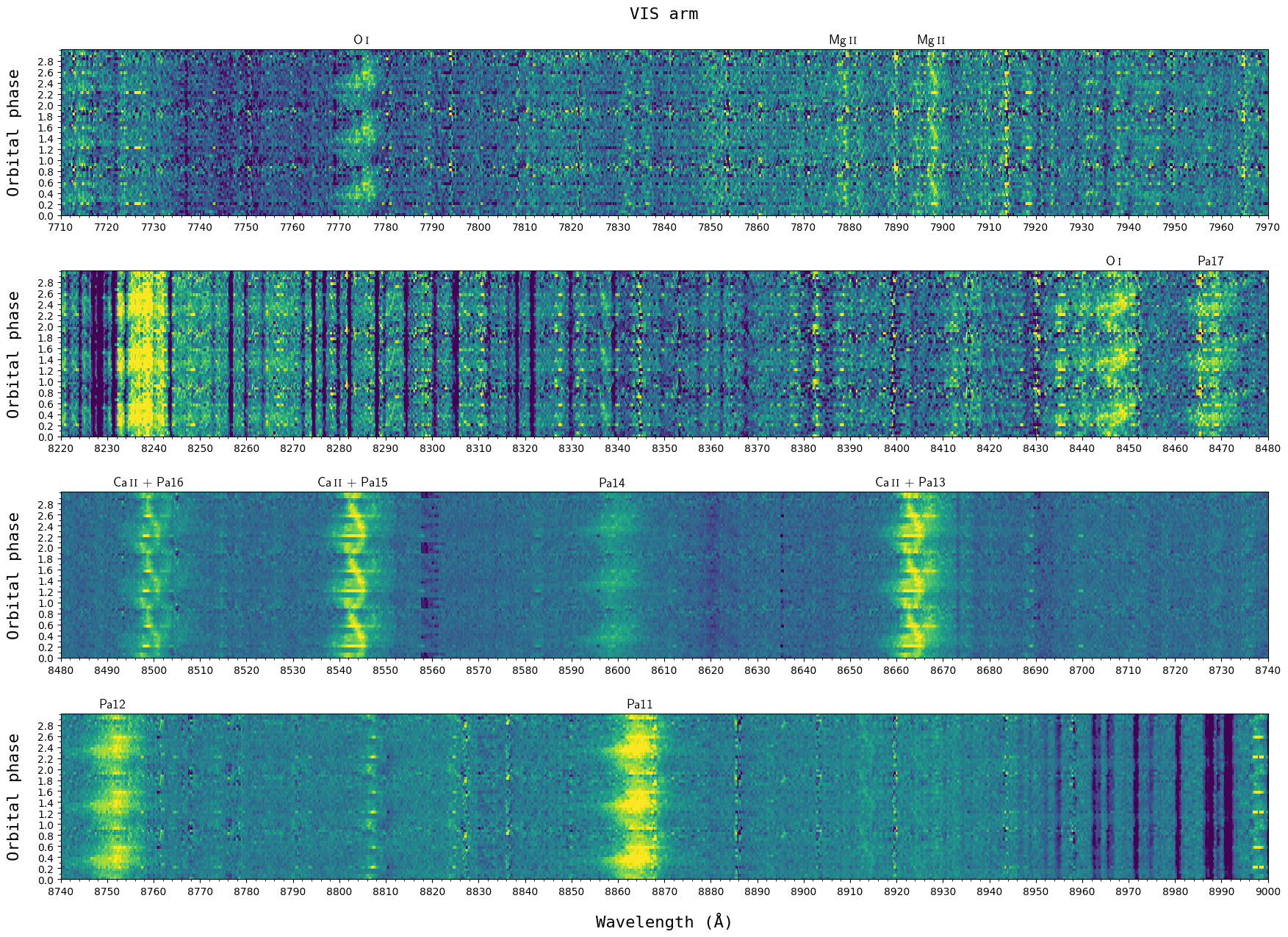}
\caption{}
\label{fig:trailed_VIS_02}
\end{center}
\end{figure}
\end{landscape}

\begin{landscape}
\begin{figure}
\begin{center}
\includegraphics[width=0.85\linewidth]{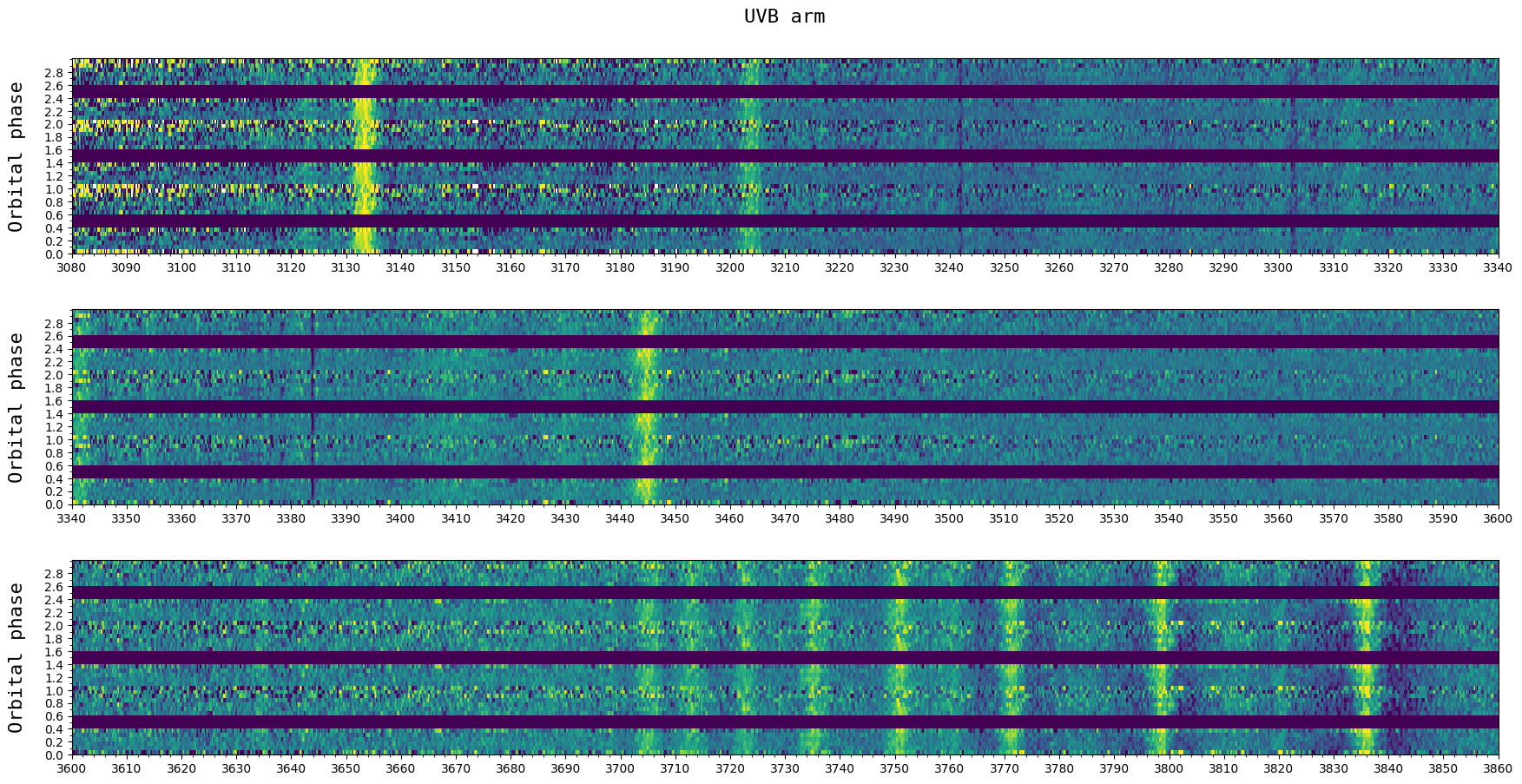}
~\par
~\par
\includegraphics[width=0.85\linewidth]{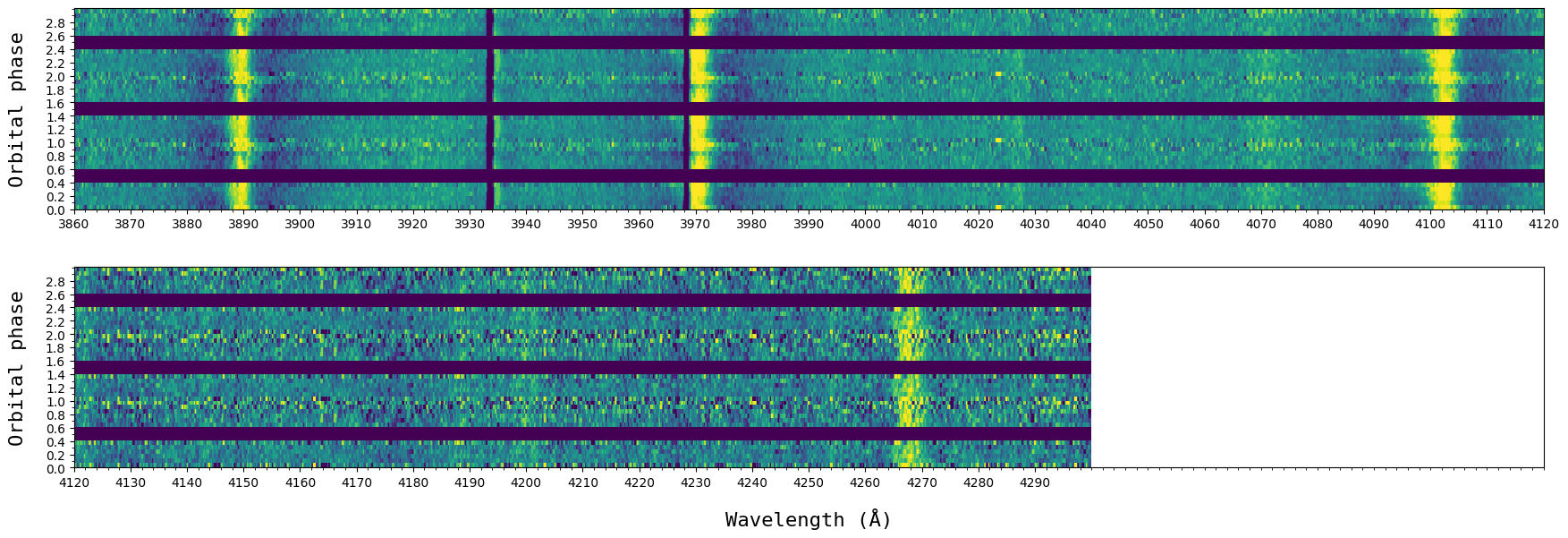}
\caption{X-shooter UVB arm trailed spectra diagrams using only the data in quiescence averaged into 15 orbital phase bins. The full cycle has been repeated twice for clarity.}
\label{fig:trailed_QUIES_UVB_01}
\end{center}
\end{figure}
\end{landscape}

\begin{landscape}
\renewcommand{\thefigure}{\thesection\arabic{figure} (cont.)}
\addtocounter{figure}{-1}
\begin{figure}
\begin{center}
\includegraphics[width=0.85\linewidth]{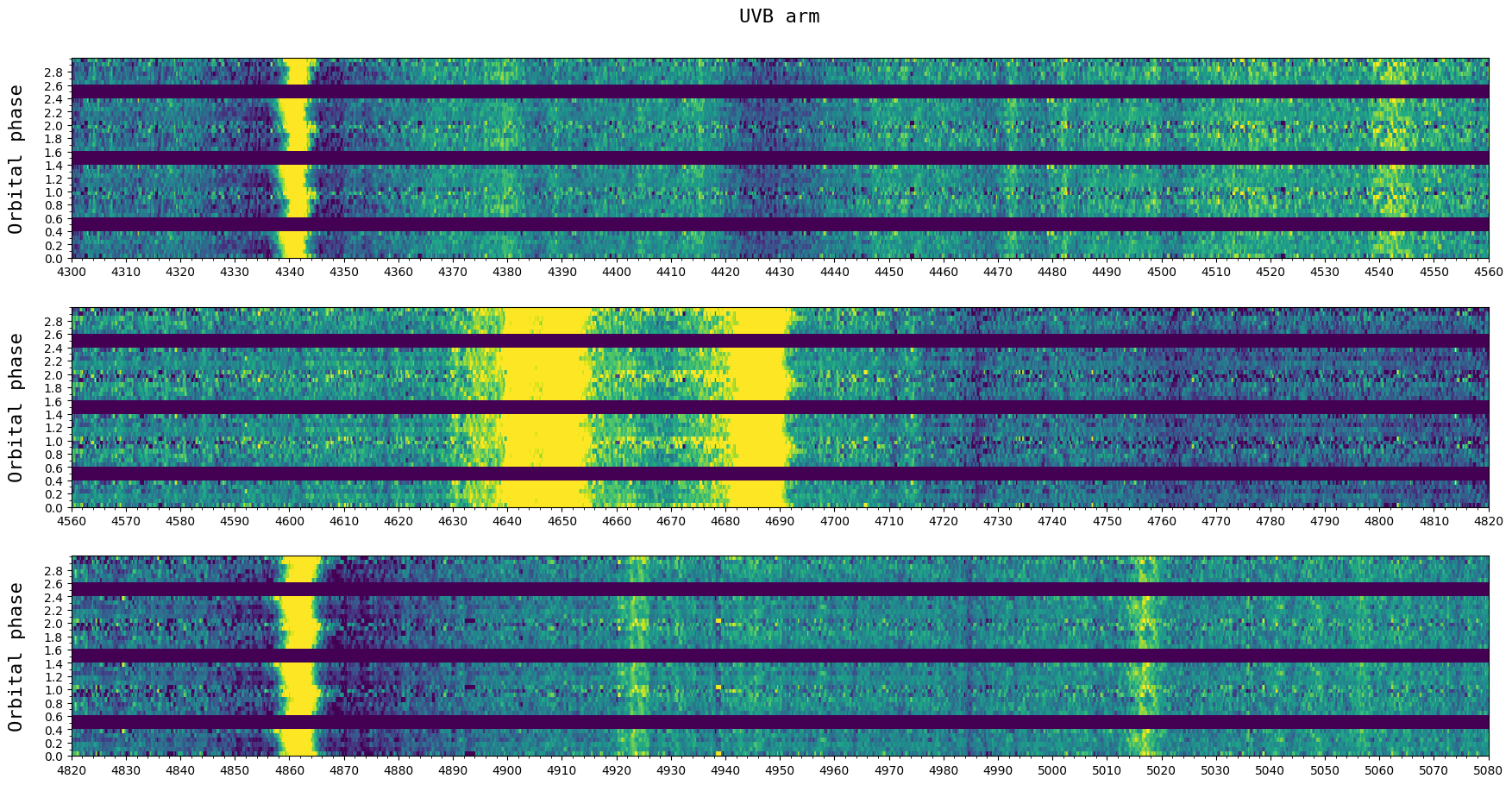}
~\par
~\par
\includegraphics[width=0.85\linewidth]{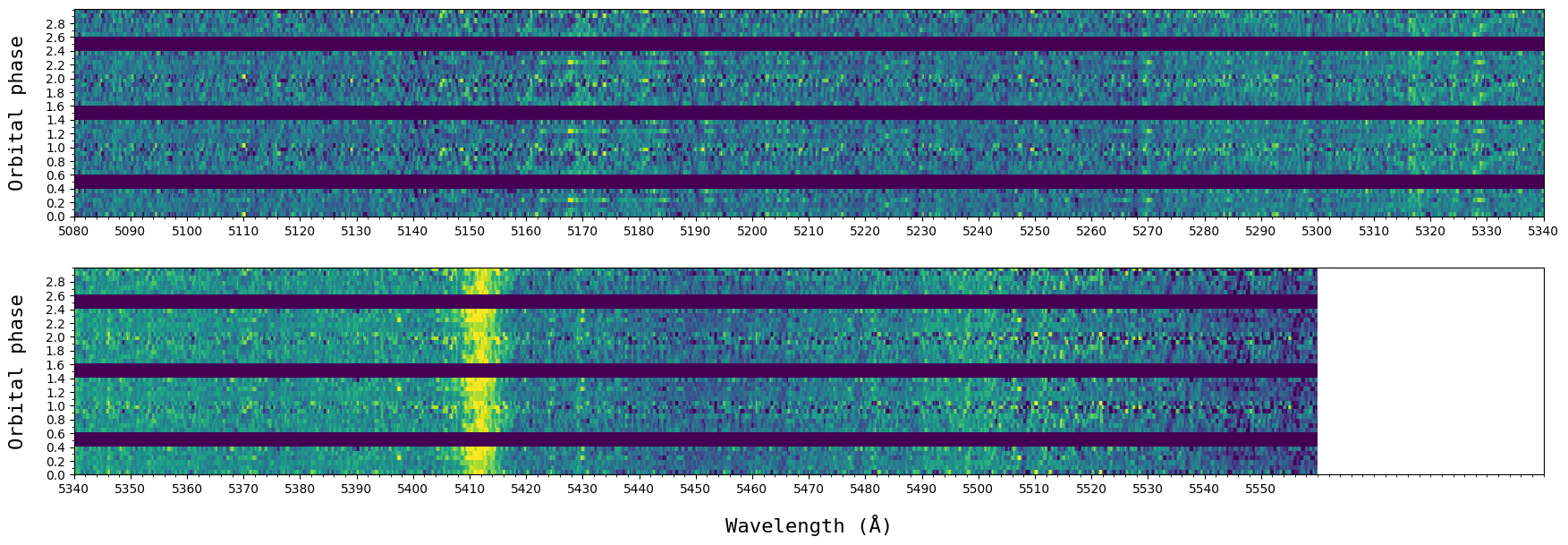}
\caption{}
\label{fig:trailed_QUIES_UVB_02}
\end{center}
\end{figure}
\end{landscape}

\begin{landscape}
\begin{figure}
\begin{center}
\includegraphics[width=0.85\linewidth]{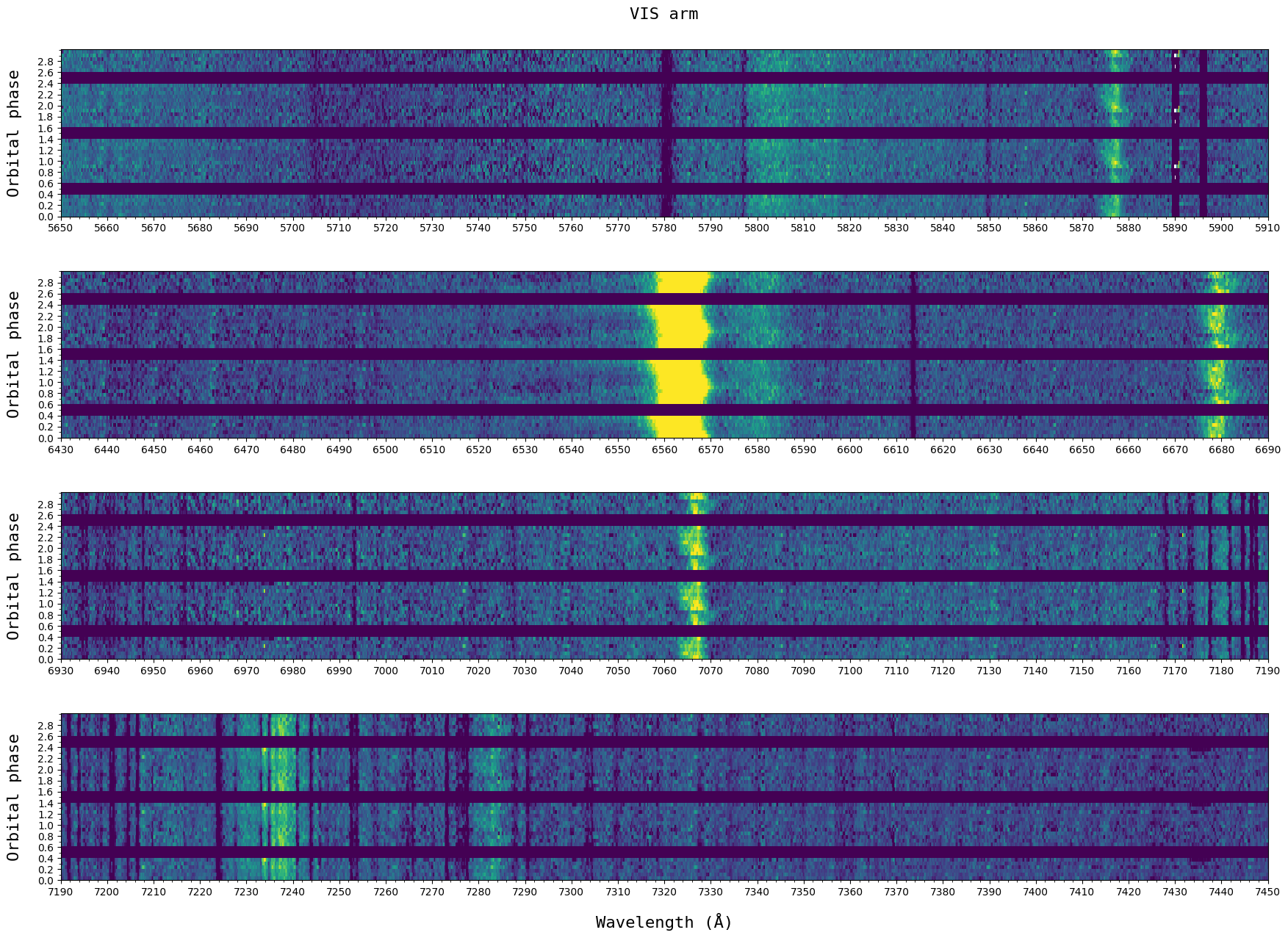}
\caption{Same as Fig.~\ref{fig:trailed_QUIES_UVB_01}, but for the VIS arm.}
\label{fig:trailed_QUIES_VIS_01}
\end{center}
\end{figure}
\end{landscape}

\begin{landscape}
\renewcommand{\thefigure}{\thesection\arabic{figure} (cont.)}
\addtocounter{figure}{-1}
\begin{figure}
\begin{center}
\includegraphics[width=0.85\linewidth]{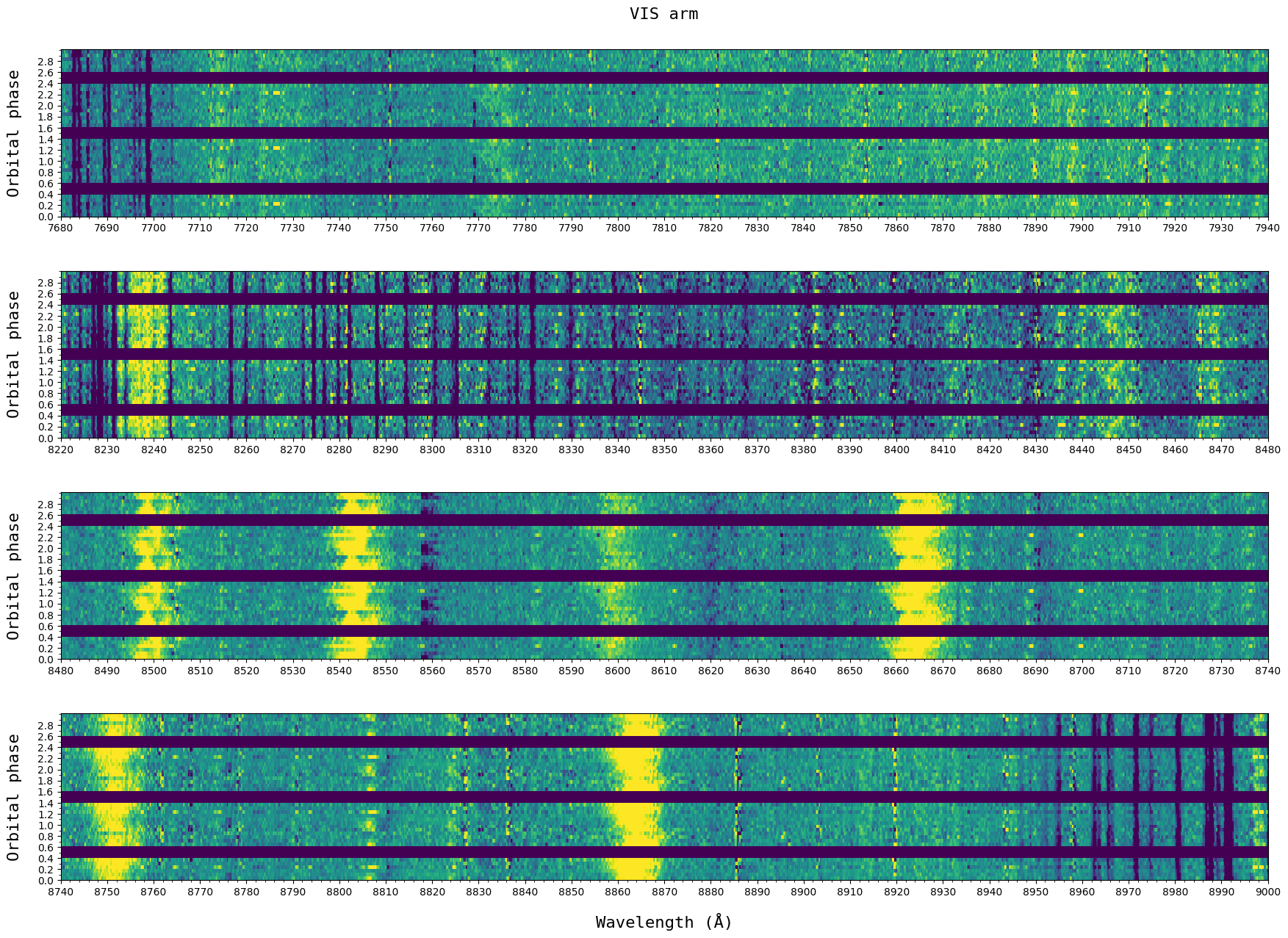}
\caption{}
\label{fig:trailed_QUIES_VIS_02}
\end{center}
\end{figure}
\end{landscape}


\bsp	
\label{lastpage}
\end{document}